\documentclass[aps,prl,nofootinbib,floatfix,twocolumn,showpacs]{revtex4}
\usepackage{graphicx}
\usepackage{color}
\usepackage{amssymb}
\usepackage{amsmath}
\usepackage{epstopdf}
\usepackage{cancel}

\usepackage{ulem}
\usepackage{slashed}
\usepackage[T1]{fontenc}
\usepackage[utf8]{inputenc}
\usepackage{lmodern}
\usepackage{amsfonts}

\newcommand{\be}{\begin{eqnarray}}
\newcommand{\ee}{\end{eqnarray}}

\newcommand{\bfq}{{\bf q}_{\perp}}

\newcommand{\bfk}{{\bf k}_{\perp}}

\begin{document}
\title{Relating transverse structure of various parton distributions}
\author{ Tanmay Maji$^1$,  Chandan Mondal$^1$, D. Chakrabarti$^{1}$ and O. V. Teryaev$^{2,3}$ }
\affiliation{ $ ^1$Department of Physics, 
Indian Institute of Technology Kanpur,
Kanpur 208016, India\\
$ ^2$Bogoliubov Laboratory of Theoretical Physics, Joint Institute of Nuclear Research, 141980 Dubna, Russia\\
$ ^3$ National Research Nuclear University "MEPhI"(Moscow Engineering Physics Institute),
Kashirskoe highway 31, 115409 Moscow, Russia}
\date{\today}

\begin{abstract}
We present the results of T-even TMDs in a light front quark-diquark model of nucleons with the wave functions constructed from the soft-wall AdS/QCD prediction.  The relations amongst TMDs are discussed. The $p_\perp$ dependence of the TMDs are compared with the $t$-dependence of the GPDs
and the relation between the respective dependencies are found.. 

\end{abstract}
\maketitle

\section{Introduction\label{intro}}
The three dimensional picture of proton is one of the most interesting and challenging tasks in particle physics. 
Recently,   there are many theoretical investigations as well as experiments (e.g.  ZEUS, COMPASS, HERMES, CLAS \cite{expts}) to understand the generalized parton distributions(GPDs) and transverse momentum dependent(TMDs) distributions  which encode  these informations.
These objects represent similar albeit non-identical transverse structure and in this paper we are aiming to relate these in a straightforward manner.  
For that purpose we use a light front quark-scalar diquark model of proton where the wave functions are constructed from AdS/QCD\cite{BT1} predictions,  and present a unified description of the TMDs and GPDs and the relations between these two different distribution functions.

The TMDs (see \cite{collins} and references therein) are required to describe the  Semi-Inclusive Deep Inelastic Scattering(SIDIS) or Drell-Yan processes, whereas GPDs (see \cite{diehl} and references therein)  are required for exclusive processes like deeply virtual Compton scattering or vector meson productions.
Three of the TMDs, $f_1(x,p_\perp),~g_{1L}(x,p_\perp),~h_1(x,p_\perp)$ are generalization of the three PDFs, the unpolarized distribution $f_1(x)$, helicity distribution function $g_{1}(x)$ and the transversity distribution $h_1(x)$.  Other TMDs do not have  simple collinear limit.
Both, TMDs and GPDs are studied in  several QCD inspired models.
In this paper, we mainly concentrate on the relations among the TMDs and the relations between TMDs and the GPDs and their moments.

TMDs have been investigated in several QCD inspired models, e.g., in a diquark spectator model \cite{Mulders,Bacchetta}, in MIT bag model\cite{bag}, in a covariant  parton model\cite{parton}.  The relations between the TMDs and  PDFs  were studied in\cite{efremov} and  the relations with GPDs were studied in detail in \cite{meissner}. 
Some relations between the TMDs and GPDs were also observed in \cite{muller}.

These relations are model dependent and it is not guaranteed that they should hold in QCD. A model independent derivation of the relations is not yet possible. Nevertheless, from phenomenological point of view, these relations may provide additional constraints on model predictions. 
Here, we demonstrate a novel relation between the TMDs
and the GPDs 
 which relates the  $t$( square of momentum transferred) dependence of GPDs with the $p_\perp^2$ dependence of TMDs.  
This may reflect the sort of Veneziano-like $s \leftrightarrow t$ duality (see \cite{Teryaev:2015tza}. Sect.4).
The $p_\perp^2$ dependence in TMDs coming form AdS/QCD wavefunction explains why the TMDs in lattice calculations   show approximate $x$ and $p_\perp^2$ factorization. The same factorization is also used in phenomenological models of TMDs.
\section{light-front diquark Model}
 In this model we assume that the incoming photon, carrying a high momentum, interacts with one of the valence quark inside the nucleon and other two valence quarks form a bound state of spin-0 (scalar diquark).  Therefore the nucleon state $|P,S\rangle$  having momentum P and spin S, can be represented as 2-particle Fock-state. In this paper we consider the scalar diquark model \cite{Gut}.

  The light-cone components of quark momentum $p$ and spectator momentum $P_X$ are
  \begin{eqnarray}
  p &\equiv & \bigg(xP^+,\frac{p^2 + |\textbf{p}_{\perp}|^2}{ xP^+},\textbf{p}_{\perp}\bigg),\\
  P_X &\equiv & \bigg((1-x)P^+,P_X^-,-\textbf{p}_{\perp}\bigg).
  \end{eqnarray}
The 2-particle Fock-state expansion for $J^z = \pm \frac{1}{2}$   are given by
 \begin{eqnarray}
  |P;\pm\rangle
& =& \sum_q \int \frac{dx~ d^2\textbf{p}_{\perp}}{2(2\pi)^3\sqrt{x(1-x)}} \bigg[ \nonumber\\
&&\psi^{q\pm}_+(x,\textbf{p}_{\perp})|+\frac{1}{2},0; xP^+,\textbf{p}_{\perp}\rangle + \nonumber \\
 && \psi^{q\pm}_-(x,\textbf{p}_{\perp})|-\frac{1}{2},0; xP^+,\textbf{p}_{\perp}\rangle\bigg],
  \end{eqnarray}
 where the $|\lambda_q,\lambda_s ; xP^+, \textbf{p}_{\perp} \rangle $ represents a two particle state with a quark of spin $\lambda_q = \pm\frac{1}{2} $, momentum $p$ and a  scalar spectator($\lambda_s=0$). The states are   normalized as: 
\be
\langle \lambda'_q,\lambda'_s; x'P^+,\textbf{p}'_{\perp} & \mid &\! \lambda_q,  \lambda_s;xP^+,\textbf{p}_{\perp}\rangle =\nonumber\\
 \prod^2_{i=1} 16 \pi^3 p^+_i& &\!\!\!\!\!\!\!  \delta(p'^+_i-p^+_i)\delta^2(\textbf{p}'_{\perp i}-\textbf{p}_{\perp i}) \delta_{\lambda'_i \lambda_i}.
\ee
 $\psi^{q\lambda_N}_{\lambda_q}$ are the light-front wave functions with nucleon helicities $\lambda_N = \pm $.
 We adopt the generic ansatz for the quark-diquark model of the valence Fock state of the nucleon LFWFs
 at a scale $\mu_0=313$~MeV as proposed in \cite{Gut} : 
\begin{eqnarray}
\psi^{q+}_+(x,\textbf{p}_{\perp})&=&\varphi^{q(1)}(x,\textbf{p}_{\perp}),\nonumber\\
\psi^{q+}_-(x,\textbf{p}_{\perp})&=&-\frac{p^1+ip^2}{xM}\varphi^{q(2)}(x,\textbf{p}_{\perp}),\nonumber\\
\psi^{q-}_+(x,\textbf{p}_{\perp})&=&\frac{p^1-ip^2}{xM}\varphi^{q(2)}(x,\textbf{p}_{\perp}),\label{lfwf}\\
\psi^{q-}_-(x,\textbf{p}_{\perp})&=& \varphi^{q(1)}(x,\textbf{p}_{\perp}),\nonumber
\end{eqnarray}
where $\varphi_q^{(1)}(x,\textbf{p}_\perp) $ and $\varphi_q^{(2)}(x,\textbf{p}_\perp) $ are the wave functions predicted by soft-wall AdS/QCD
\be
\varphi^{q(i)}(x,\textbf{p}_\perp)&=&N_q^{(i)}\frac{4\pi}{\kappa}\sqrt{\frac{\log(1/x)}{1-x}}x^{a_q^{(i)}}(1-x)^{b_q^{(i)}}\nonumber\\
&&\exp\bigg[-\frac{\textbf{p}_\perp^2}{2\kappa^2}\frac{\log(1/x)}{(1-x)^2}\bigg],\label{adswf}
\ee
where $\kappa=0.4$ GeV \cite{CM1} is the AdS/QCD scale parameter. 
The  values of the parameters $a^{(i)}_q$ and $b^{(i)}_q$ and the constants $N^{(i)}_q$ were fixed by fitting the nucleon form factors. For $\kappa=0.4$ GeV, the parameters are\cite{Chi} 
$a^{(1)}_u=0.02,~a^{(2)}_u= 1.05,~b^{(1)}_u= 0.022,~ b^{(2)}_u= -0.15, N^{(1)}_u=  2.055,  N^{(2)}_u=1.322,~  a^{(1)}_d=0.1,~ a^{(2)}_d= 1.07, ~b^{(1)}_d= 0.38, ~b^{(2)}_d=-0.2,~ N^{(1)}_d=  1.7618,~ \&~ N^{(2)}_d=-2.4827.$  For $a_q^{(i)}=b_q^{(i)} =0$, the wave functions reduce to the AdS/QCD prediction\cite{BT}.
 
A   transversely polarized  nucleon  with polarization $\hat{S}_T=(\cos\phi_S , \sin\phi_S)$ in the transverse plane can be written as 
\begin{equation}
\mid P;S_T\rangle = \frac{1}{\sqrt{2}}\bigg(|P;+\rangle + e^{i\phi_S} |P; - \rangle \bigg)
\end{equation}
 Without loss of generality, we choose the  nucleon polarization along $x$ axis i.e.,  $\phi_S = 0 $.

\section{TMDs}
The unintegrated quark-quark correlator for polarized SIDIS is defined  as 
\be
\Phi^{q[\Gamma]}(x,\textbf{p}_{\perp};S)&=&\frac{1}{4}\int \frac{dz^-}{(2\pi)} \frac{d^2z_T}{(2\pi)^2} e^{ip.z} \nonumber\\
&&\!\!\!\!\! \langle P; S|\overline{\psi}^q(0)\Gamma \mathcal{W}_{[0,z]} \psi^q(z) |P;S\rangle,
\ee
for a quark q. The summations over the color indices of quarks are implied. Here $p$ is the momentum of the struck quark inside the nucleon having momentum P, helicity S and $x ~(x=p^+/P^+)$ is the longitudinal momentum fraction carried by struck quark. 
We choose
 the lightcone gauge $A^+=0$ and
 a frame where the nucleon momentum and quark momentum are $ P\equiv (P^+,\frac{M^2}{P^+},\textbf{0} ),~ q\equiv (x_B P^+, \frac{Q^2}{x_BP^+},\textbf{0})$ respectively, $x_B= \frac{Q^2}{2P.q}$ is the Bjorken Scaling with $Q^2 = -q^2$. The nucleon with helicity $\lambda$ has  spin components $S^+ = \lambda \frac{P^+}{M},~ S^- = \lambda\frac{P^-}{M},$ and $ S_T $.  
In leading twist, the TMDs are defined as
\begin{eqnarray}
\Phi^{q[\gamma^+]}(x,\textbf{p}_{\perp};S)&=& f_1^q(x,\textbf{p}_{\perp}^2) - \frac{\epsilon^{ij}_Tp^i_\perp S^j_T}{M}f^{\perp q}_{1T}(x,\textbf{p}_{\perp}^2),\nonumber\\
\Phi^{q[\gamma^+ \gamma^5]}(x,\textbf{p}_{\perp};S) &=&  \lambda g_{1L}^q(x,\textbf{p}_{\perp}^2) + \frac{\textbf{p}_{\perp}.\textbf{S}_T}{M} g^q_{1T}(x,\textbf{p}_{\perp}^2),\nonumber\\
\Phi^{q[i \sigma^{j +}\gamma^5]}(x,\textbf{p}_{\perp};S)& = & S^j_T h_1^q(x,\textbf{p}_{\perp}^2) + \lambda\frac{p^j_\perp}{M}h^{\perp q}_{1L}(x,\textbf{p}_{\perp}^2)\nonumber\\
&+& \frac{2 p^j_\perp \textbf{p}_{\perp}.\textbf{S}_T - S^j_T \textbf{p}^2_{\perp}}{2M^2} h^{\perp q}_{1T}(x,\textbf{p}_{\perp}^2) \nonumber\\
&-& \frac{\epsilon_T^{ij}p^i_{\perp}}{M}h^{\perp q}_1(x,\textbf{p}_{\perp}^2).
\end{eqnarray}
The $p_\perp$ integrated function of $f_1^q(x,p_\perp^2)$ gives the unpolarized distribution $f^{q}_1(x)$ and that of $g_{1L}^q(x,p_\perp^2)$ ($=g_1^q(x,p_\perp^2)$) gives the helicity distribution $g^{q}_1(x)$. The $h^{q}_1 (x,\textbf{p}_{\perp}^2)$ is the transversity TMD which gives the transversity distribution $h^{q}_1 (x)$(integrating over $ p_\perp $):
\begin{eqnarray}
h^{q}_1 (x,\textbf{p}_{\perp}^2) &=& h^{q}_{1T} (x,\textbf{p}_{\perp}^2) + \frac{\textbf{p}^2_{\perp}}{2M^2}h^{\perp q}_{1T} (x,\textbf{p}_{\perp}^2)\label{h1q}\\
h^{q}_1 (x) &=& \int d^2p_\perp h^{q}_1 (x,\textbf{p}_{\perp}^2)
\end{eqnarray} 

The transverse momentum dependent parton distributions, in terms of Light-Front Wave Functions(LFWFs) $\psi^{q \lambda_N} _{\lambda_q}$ take the following forms
\begin{widetext}
\be
{f}^{q  }_1(x,\textbf{p}_{\perp}^2)&=&\frac{1}{16\pi^3}\bigg[|\psi ^{q+}_+(x,\textbf{p}_{\perp})|^2+|\psi ^{q+}_-(x,\textbf{p}_{\perp})|^2\bigg],\\ 
{g}^{q  }_{1L}(x,\textbf{p}_{\perp}^2)&=&\frac{1}{16\pi^3}\bigg[|\psi ^{q+}_+(x,\textbf{p}_{\perp})|^2 - |\psi ^{q+}_-(x,\textbf{p}_{\perp})|^2\bigg],\\
\frac{\textbf{p}_{\perp}.\textbf{S}_T}{M}~ {g}^{q  }_{1T}(x,\textbf{p}_{\perp}^2)&=& - \frac{1}{16\pi^3}\bigg[\psi ^{q+}_+(x,\textbf{p}_{\perp})\psi^{q+\dagger}_-(x,\textbf{p}_{\perp})
+{\psi ^{q+}_-}(x,\textbf{p}_{\perp}){\psi^{q+}_+}^\dagger(x,\textbf{p}_{\perp})\bigg],\\
\frac{\textbf{p}_{\perp}.\textbf{s}_{qT}}{M} ~{h}^{q\perp  }_{1L}(x,\textbf{p}_{\perp}^2)&=&\frac{1}{16\pi^3}\bigg[\psi ^{q+\dagger}_+(x,\textbf{p}_{\perp})\psi ^{q+}_-(x,\textbf{p}_{\perp}) + \psi ^{q+\dagger}_-(x,\textbf{p}_{\perp})\psi ^{q+}_+(x,\textbf{p}_{\perp})\bigg], \\
\textbf{S}_T . \textbf{s}_{qT}~ {h}^{q  }_{1T}(x,\textbf{p}_{\perp}^2) ~+&& \frac{\textbf{p}_{\perp} . \textbf{S}_{T}~\textbf{p}_{\perp} . \textbf{s}_{qT}}{M^2} {h}^{\perp q  }_{1T}(x,\textbf{p}_{\perp}^2)
\nonumber\\  ~~
&=&\frac{1}{16\pi^3}\bigg[|\psi ^{q+}_+(x,\textbf{p}_{\perp})|^2 - \frac{1}{2}(\psi ^{q+\dagger}_-(x,\textbf{p}_{\perp}))^2 - \frac{1}{2}(\psi ^{q+}_-(x,\textbf{p}_{\perp}))^2\bigg].
\ee
\end{widetext}
Using the light-front wave functions from Eq. (\ref{lfwf}) and Eq.(\ref{adswf}), the  explicit expressions for the TMDs can be written as:
\be
{f}^{q  }_1(x,\textbf{p}^2_{\perp})&=&\frac{\log(1/x)}{\pi\kappa^2}\exp\bigg[-\frac{\textbf{p}_{\perp}^2\log(1/x)}{\kappa^2(1-x)^2}\bigg]\nonumber\\
&&\bigg(F_1(x) +\frac{\textbf{p}^2_{\perp}}{M^2} F_2(x)\bigg),\label{TMDs}\\
{g}^{q  }_{1L}(x,\textbf{p}^2_{\perp})&=&\frac{\log(1/x)}{\pi\kappa^2}\exp\bigg[-\frac{\textbf{p}_{\perp}^2\log(1/x)}{\kappa^2(1-x)^2}\bigg]\nonumber\\
&&\bigg(F_1(x)-
 \frac{\textbf{p}^2_{\perp}}{M^2} F_2(x)\bigg),\label{g1L} \nonumber\\
{h}^{q  }_1(x,\textbf{p}^2_{\perp}) &=& \frac{\log(1/x)}{\pi\kappa^2}\exp\bigg[-\frac{\textbf{p}_{\perp}^2\log(1/x)}{\kappa^2(1-x)^2}\bigg]F_1(x), \label{h1}\nonumber\\
{g}^{q  }_{1T}(x,\textbf{p}^2_{\perp})&=&\frac{2\log(1/x)}{\pi\kappa^2}\exp\bigg[-\frac{\textbf{p}_{\perp}^2\log(1/x)}{\kappa^2(1-x)^2}\bigg]F_3(x),\nonumber\\
{h}^{q\perp  }_{1L}(x,\textbf{p}^2_{\perp})&\!\!=&\!\!\!-\frac{2\log(1/x)}{\pi\kappa^2}\exp\bigg[-\frac{\textbf{p}_{\perp}^2\log(1/x)}{\kappa^2(1-x)^2}\bigg] F_3(x),\nonumber\\
{h}^{q  }_{1T}(x,\textbf{p}^2_{\perp})&=& \frac{\log(1/x)}{\pi\kappa^2}\exp\bigg[-\frac{\textbf{p}_{\perp}^2\log(1/x)}{\kappa^2(1-x)^2}\bigg]\nonumber\\
&&~\bigg(F_1(x)+ \frac{\textbf{p}^2_{\perp}}{M^2}  F_2(x)\bigg),\nonumber\\
{h}^{q\perp  }_{1T}(x,\textbf{p}^2_{\perp})&\!\!=&\!\! - \frac{2\log(1/x)}{\pi\kappa^2}\exp\bigg[-\frac{\textbf{p}_{\perp}^2\log(1/x)}{\kappa^2(1-x)^2}\bigg]F_2(x),\nonumber
\ee
where 
\be
F_1(x)&=&|N^{(1)}_q|^2 x^{2a^{(1)}_q}(1-x)^{2b^{(1)}_q-1}, \nonumber\\
F_2(x)&=&|N^{(2)}_q|^2 x^{2a^{(2)}_q -2}(1-x)^{2b^{(2)}_q-1},\label{Fx}\\
F_3(x)&=&N^{(1)}_q N^{(2)}_q x^{a^{(1)}_q +a^{(2)}_q-1}(1-x)^{b^{(1)}_q+b^{(2)}_q-1}.\nonumber
\ee
$SU(6)$ spin-flavor symmetry requires that all the polarized TMDs to be scaled by the flavor factor $P_q$ where $P_u=\frac{4}{3}$ and $P_d=-\frac{1}{3}$ \cite{Karl}.
Since in the quark-diquark model, the sea quarks are ignored, here we have the valence TMDs. But it should be noted that  the AdS/QCD wavefunctions adopted here  are not ``purely valence" wavefunctions but effective valence wavefunctions which encode  aspects of nonperturbative dynamics \cite{BT} and cannot be obtained in a model with only valence quarks, hence we expect the distribution functions also encode informations beyond valence quarks.

\subsection{$p_\perp$- integrated distributions:}
The parton distribution functions(PDFs) in this quark-diquark model and their scale evolutions have been discussed in detail in \cite{Gut} and showed to agree with a global fit for both $u$ and $d$ quarks.
The $\textbf{p}_{\perp}$ integrated distribution functions  at the initial scale $\mu_0=313$ MeV  are
\be
f^{q  }_1(x)&=& F_1(x)(1-x)^2+\nonumber\\
&&~F_2(x)(1-x)^4\frac{\kappa^2}{M^2\ln(1/x)}, \\
g^{q  }_{1}(x)&=& F_1(x)(1-x)^2 ~ - \nonumber\\
&&~F_2(x)(1-x)^4\frac{\kappa^2}{M^2\ln(1/x)}, \\
h^{q  }_1(x) &=&F_1(x)(1-x)^2,\\
g^{q  }_{1T}(x)&=& 2 F_3(x) (1-x)^2,\\
h^{\perp q  }_{1L}(x)&=& - 2 F_3(x) (1-x)^2,\\
h^{q  }_{1T}(x)&=& F_1(x)(1-x)^2+\nonumber\\
&&~F_2(x)(1-x)^4\frac{\kappa^2}{M^2\ln(1/x)} ,\\
h^{\perp q  }_{1T}(x)&=& - 2 F_2(x) (1-x)^2.
\ee

\subsection{TMD relations}
The relations satisfied by the TMDs  in the light-front diquark model(Eqs.(\ref{TMDs}-\ref{Fx})) are the  following 
\be 
| {h}^{q  }_1(x,\textbf{p}^2_{\perp})| &=& \frac{1}{2} \bigg[ {f}^{q  }_1(x,\textbf{p}^2_{\perp}) + {g}^{q  }_{1L}(x,\textbf{p}^2_{\perp}) \bigg],\label{soffer_bound}\\ 
{g}^{q  }_{1T}(x,\textbf{p}^2_{\perp}) &=&  - {h}^{q\perp  }_{1L}(x,\textbf{p}^2_{\perp}),\\ 
{h}^{q  }_{1T}(x,\textbf{p}^2_{\perp}) &=& {f}^{q  }_1(x,\textbf{p}^2_{\perp}),\\
\frac{\textbf{p}^2_\perp}{2 M^2}{h}^{q\perp  }_{1T}(x,\textbf{p}^2_{\perp}) &=& {g}^{q  }_{1L}(x,\textbf{p}^2_{\perp})- {h}^{q  }_1(x,\textbf{p}^2_{\perp}),\\
\frac{\textbf{p}^2_\perp}{2 M^2}{h}^{q\perp  }_{1T}(x,\textbf{p}^2_{\perp}) &=& \frac{1}{2}\Big[ {g}^{q  }_{1L}(x,\textbf{p}^2_{\perp}) - {f}^{q  }_1(x,\textbf{p}^2_{\perp})\Big] . 
\ee 
The Eq.(\ref{soffer_bound}) satisfies the saturation condition of Soffer bound \cite{soffer}. 
The leading twist TMDs in the diquark model also satisfy the inequality relations which are valid in QCD and all models\cite{bounds,bag}:
\be
{f}^{q  }_1(x,\textbf{p}^2_{\perp}) &\geq & 0 ,\\
\mid {g}^{q  }_{1L}(x,\textbf{p}^2_{\perp})\mid & \leq & \mid {f}^{q  }_1(x,\textbf{p}^2_{\perp})\mid,\\
\mid {h}^{q  }_1(x,\textbf{p}^2_{\perp})\mid & \leq &\mid {f}^{q  }_1(x,\textbf{p}^2_{\perp})\mid.
\ee
From Eqs. (\ref{h1q}),(\ref{TMDs}) and (\ref{Fx}),  it is easy to see that 
\be \mid {h}_1^q(x,p_\perp^2)\mid >\mid{ g}_{1L}^q(x,p_\perp^2)\mid, \ee
 which was also observed in parton model \cite{parton} 
 and is the generalization of the relation between the tensor charge ($g_T^q$) and axial charge ($g_A^q$)
 \be \mid g_T^q\mid > \mid g_A^q\mid,\ee 
found in many models and lattice QCD(see \cite{parton} and references therein).
A non-linear relation is satisfied as
\be
{h}^{q  }_1(x,\textbf{p}^2_{\perp})~ {h}^{q\perp  }_{1T}(x,\textbf{p}^2_{\perp}) =- \frac{1}{2} \bigg[ {h}^{q\perp  }_{1L}(x,\textbf{p}^2_{\perp})\bigg]^2
\ee
The above relations are consistent with the relations found in other models like \cite{bag} and are proved to be generic for scalar diquark models\cite {Lorce:2011zta}. All the  relations listed above are independent of the parameters in our model.

\subsection{Comparison with lattice QCD}
\begin{figure*}[]
\begin{minipage}[c]{.98\textwidth}
\small{(1a)}\includegraphics[width=7.5cm,clip]{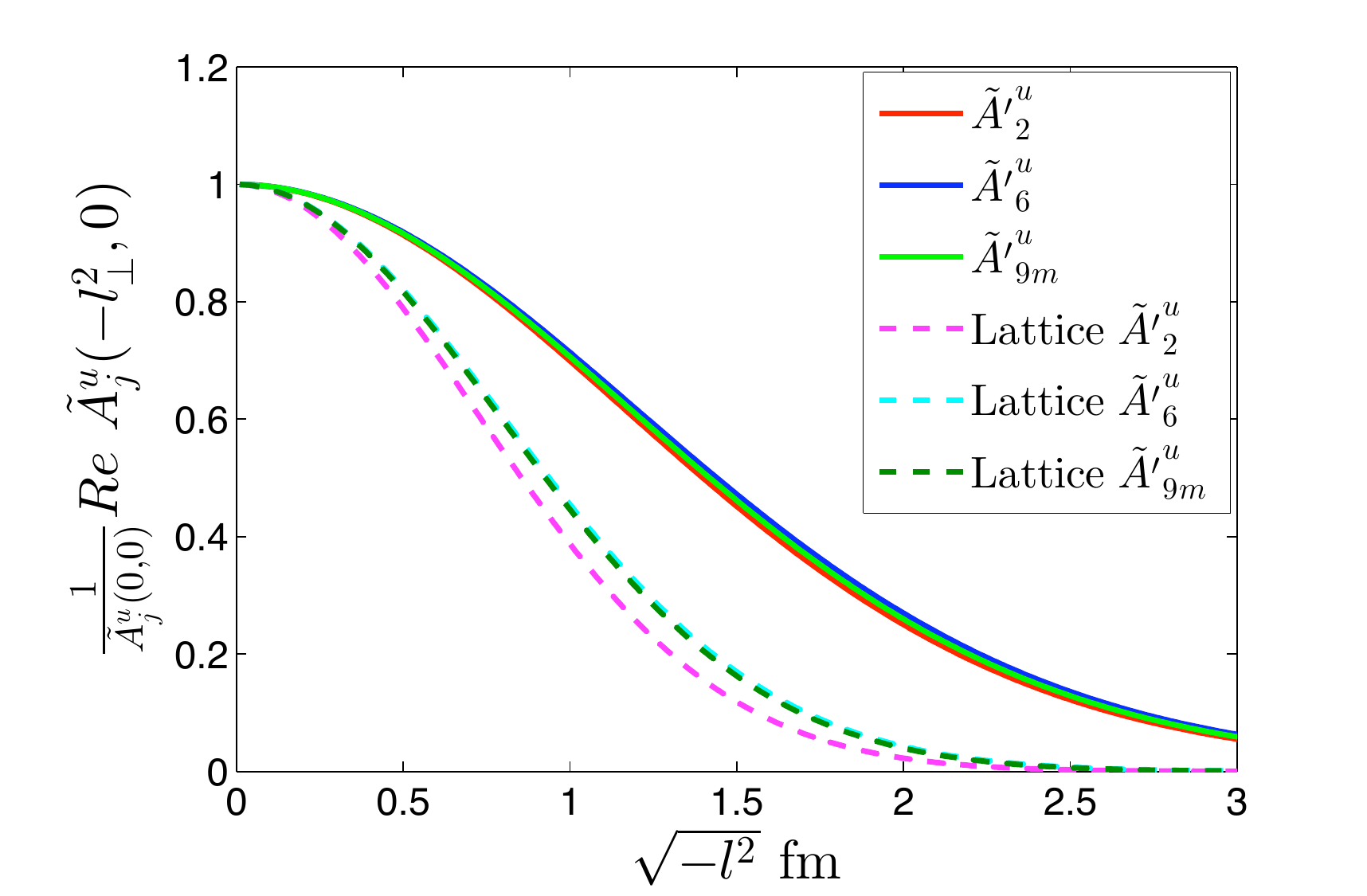}
\hspace{0.1cm}%
\small{(1b)}\includegraphics[width=7.5cm,clip]{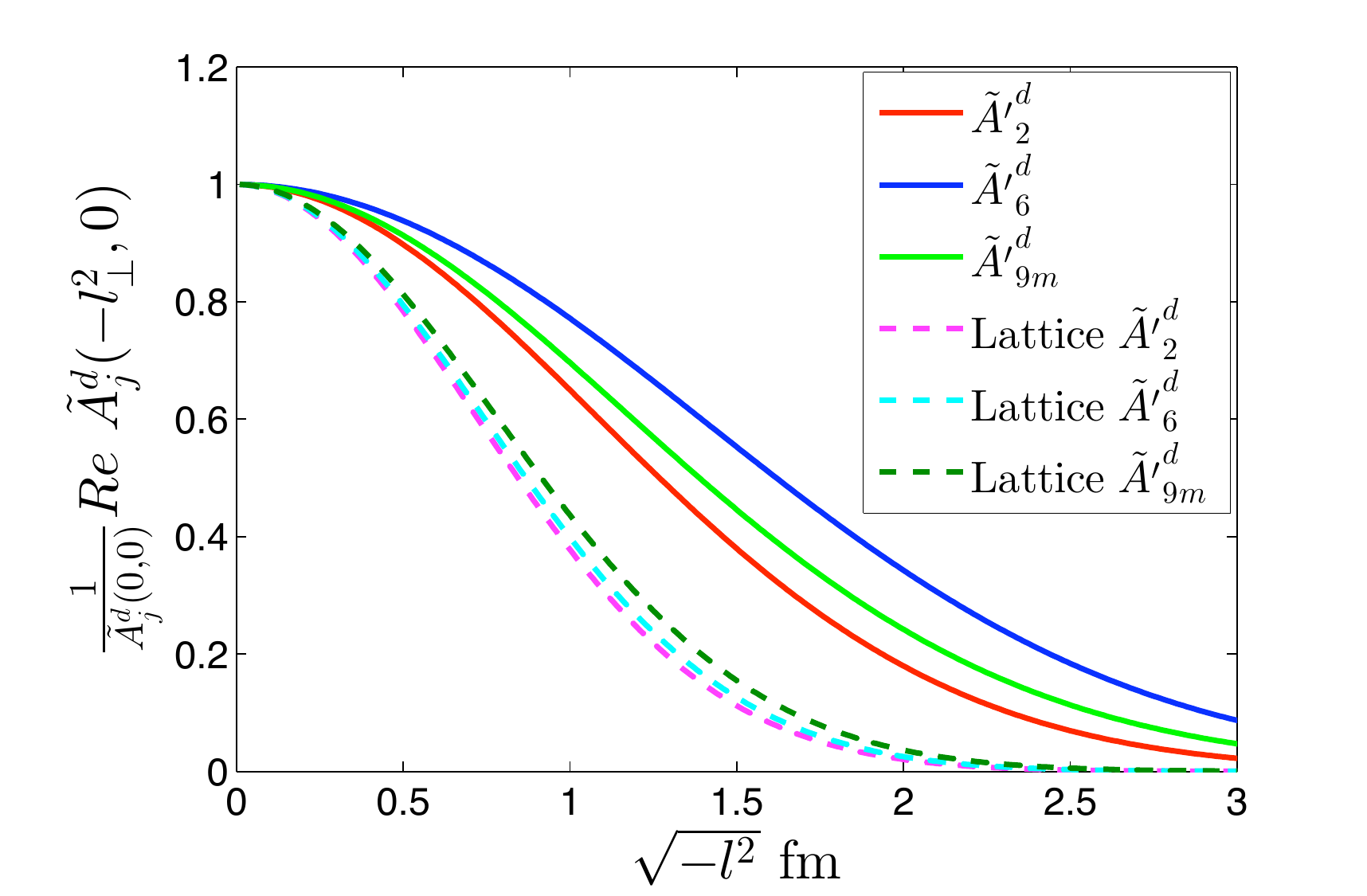}
\end{minipage}
\caption{Comparison of  ${A}_i(-\textbf{l}^2_\perp,0)$ in lattice QCD and  AdS/QCD (a) for  u and (b) for  d quark. The doted lines are the Gaussian fits of the lattice data taken from \cite{Hagler}, solid lines are the AdS/QCD results. In the plots, $\tilde{A}_i^{\prime q}= \frac{Re \tilde{A}_i^q(-l_\perp^2,0)}{A_i^q(0,0)}$.  \label{lattice_comparison}}
\end{figure*}

In lattice QCD the hadronic matrix elements can be parametrized by the invariant amplitudes $\tilde{A}_i(l^2;P,S)$ as \cite{Hagler}
\be
\tilde{{\Phi}}^{[\gamma^+]} &=& 4 P^+ \tilde{A}_2 + 4iM^2 l^+ \tilde{A}_3,\\ \label{latticeV}
\tilde{{\Phi}}^{[\gamma^+\gamma^5]} =&&\!\!\! \!\!\!\!\!\!\!-4 M S\mu \tilde{A}_6 - 4iM P^+l.S \tilde{A}_7+4M^3 l^+ l.S \tilde{A}_8,\nonumber\\
\tilde{{\Phi}}^{[i\sigma^{j+}\gamma^5]} &=& 4 S^{[j}P^{+]} \tilde{A}_{9m} + 4 i M^2 S^{[j}l^{+]} \tilde{A}_{10} \nonumber\\
&-& 2 M^2 [2 l.S l^{[j}P^{+]} - l^2 S^{[j}P^{+]}]\tilde{A}_{11}.\label{latticeT}
\ee
Where $\tilde{\Phi}^{[\Gamma]}$ is defined as 
\be
\!\!\!\Phi^{[\Gamma]}(p;s)=\!\frac{1}{4}\int\!\! \frac{d^2l_\perp dl^-}{(2\pi)^3}e^{-{i\over 2}p^+l^-+ip_\perp\cdot l_\perp}\tilde{\Phi}^{[\Gamma]}(l,p;s).
\ee
In the limit $l.P = 0$ for $l^+ = 0 $, the amplitudes in the coordinate space  can be expressed in terms of the 1st-moment of TMDs, e.g,
\be 
2 \tilde{A}_2(-\textbf{l}^2_\perp,0) &=& \int d^2p_\perp e^{-i \textbf{p}_\perp.\textbf{l}_\perp} f^{(1)}_1(x,\textbf{p}^2_\perp),\\ \label{A2}
2 \tilde{A}_6(-\textbf{l}^2_\perp,0) &=&\!\!\! - \int d^2p_\perp e^{-i \textbf{p}_\perp.\textbf{l}_\perp} g^{(1)}_{1L}(x,\textbf{p}^2_\perp),\\ \label{A6}
2 \tilde{A}_{9m}(-\textbf{l}^2_\perp,0) &=& \int d^2p_\perp e^{-i \textbf{p}_\perp.\textbf{l}_\perp} h^{(1)}_1(x,\textbf{p}^2_\perp). \label{A9m}
\ee 
The superscript $(1)$ on the TMDs indicates the first moment.
In Fig.\ref{lattice_comparison}, we have compared our model  with the corresponding lattice results\cite{Hagler}. It is interesting to note that the above three lattice moments are almost identical when normalized to one at $l_\perp^2=0$.  The diquark model also show similar behavior.

\section{quark densities}
\begin{figure*}[]
\begin{minipage}[c]{0.98\textwidth}
\small{(a)}\includegraphics[width=7.5cm,height=5.5cm,clip]{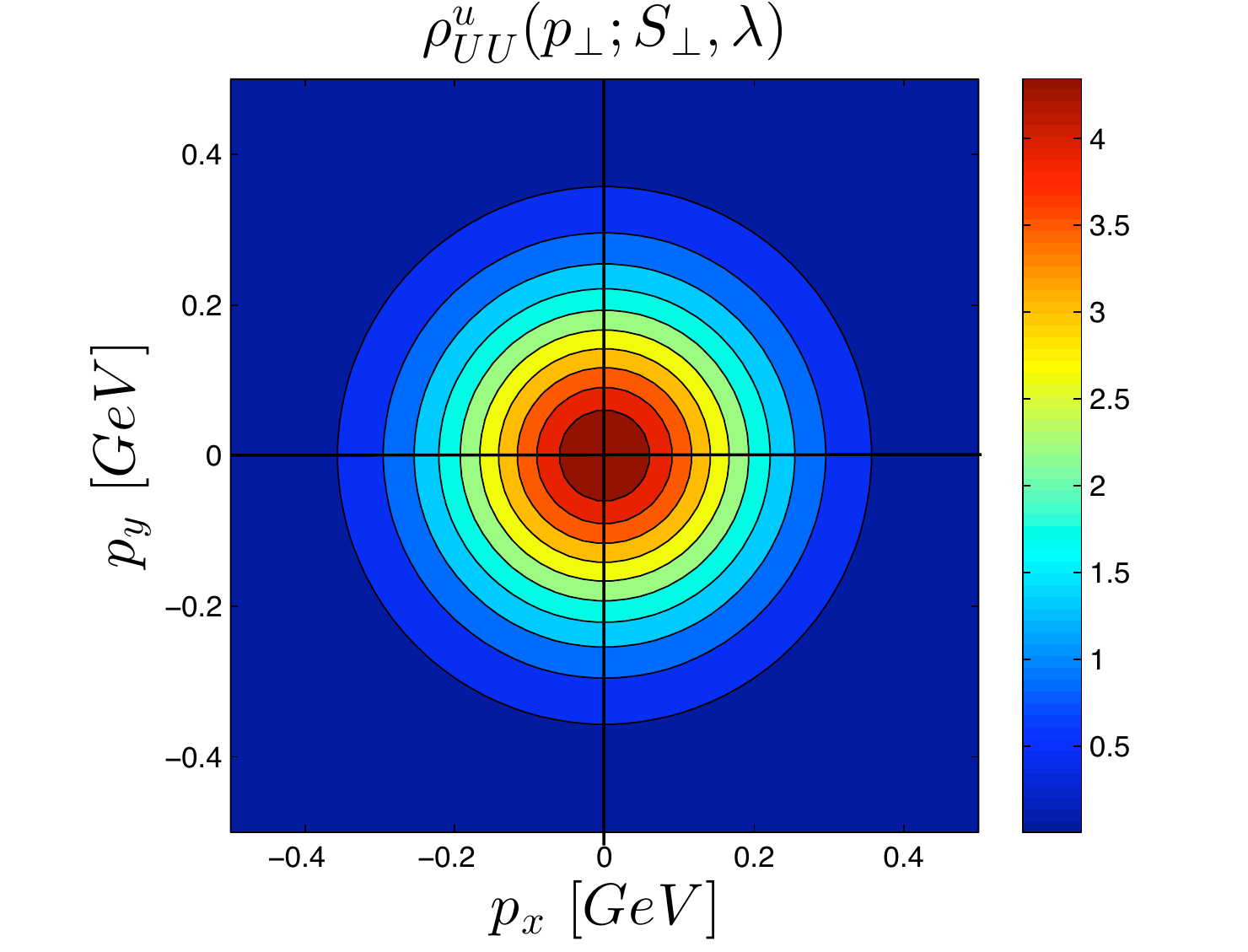}
\hspace{0.1cm}%
\small{b)}\includegraphics[width=7.5cm,height=5.5cm,clip]{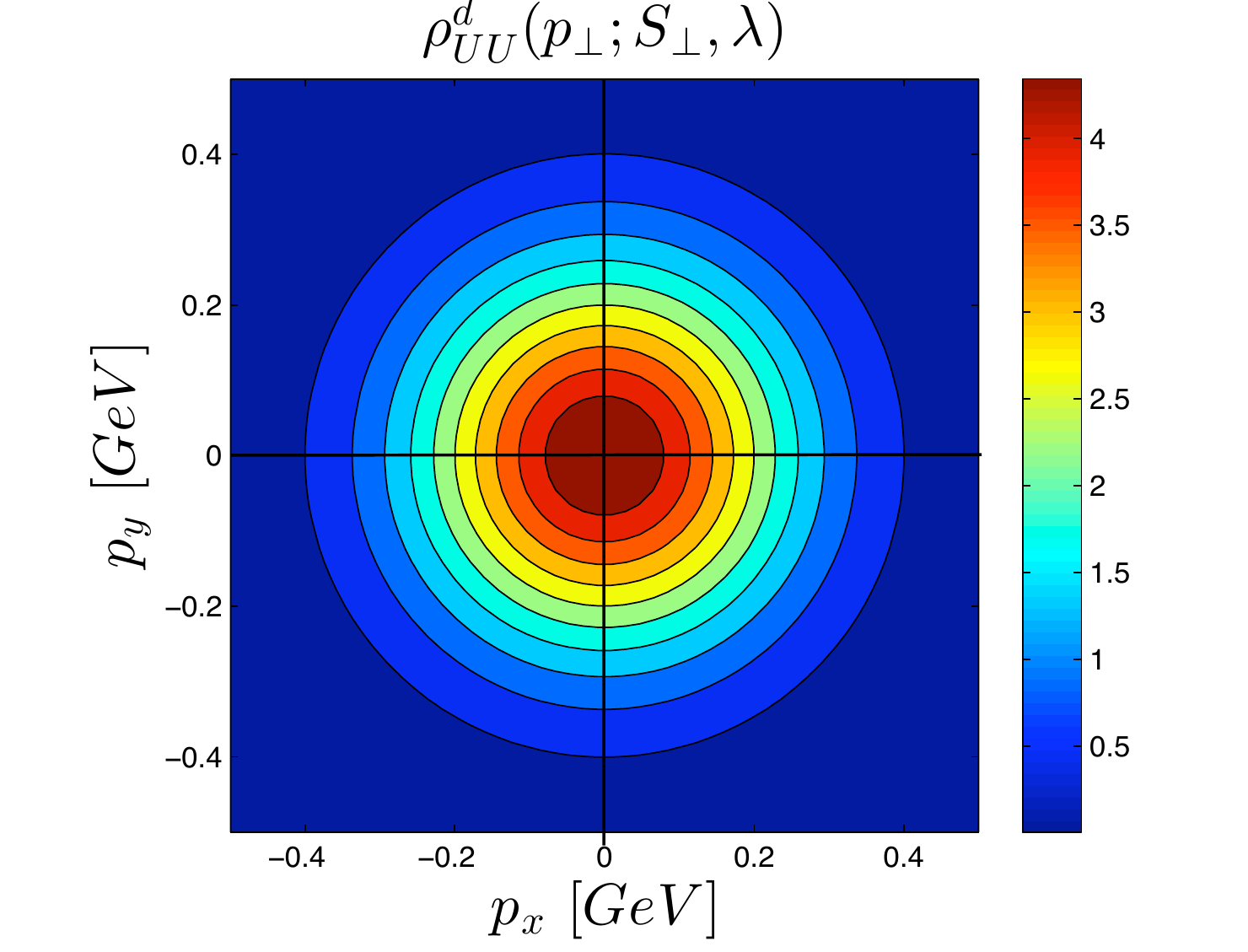}
\small{(c)}\includegraphics[width=7.5cm,height=5.5cm,clip]{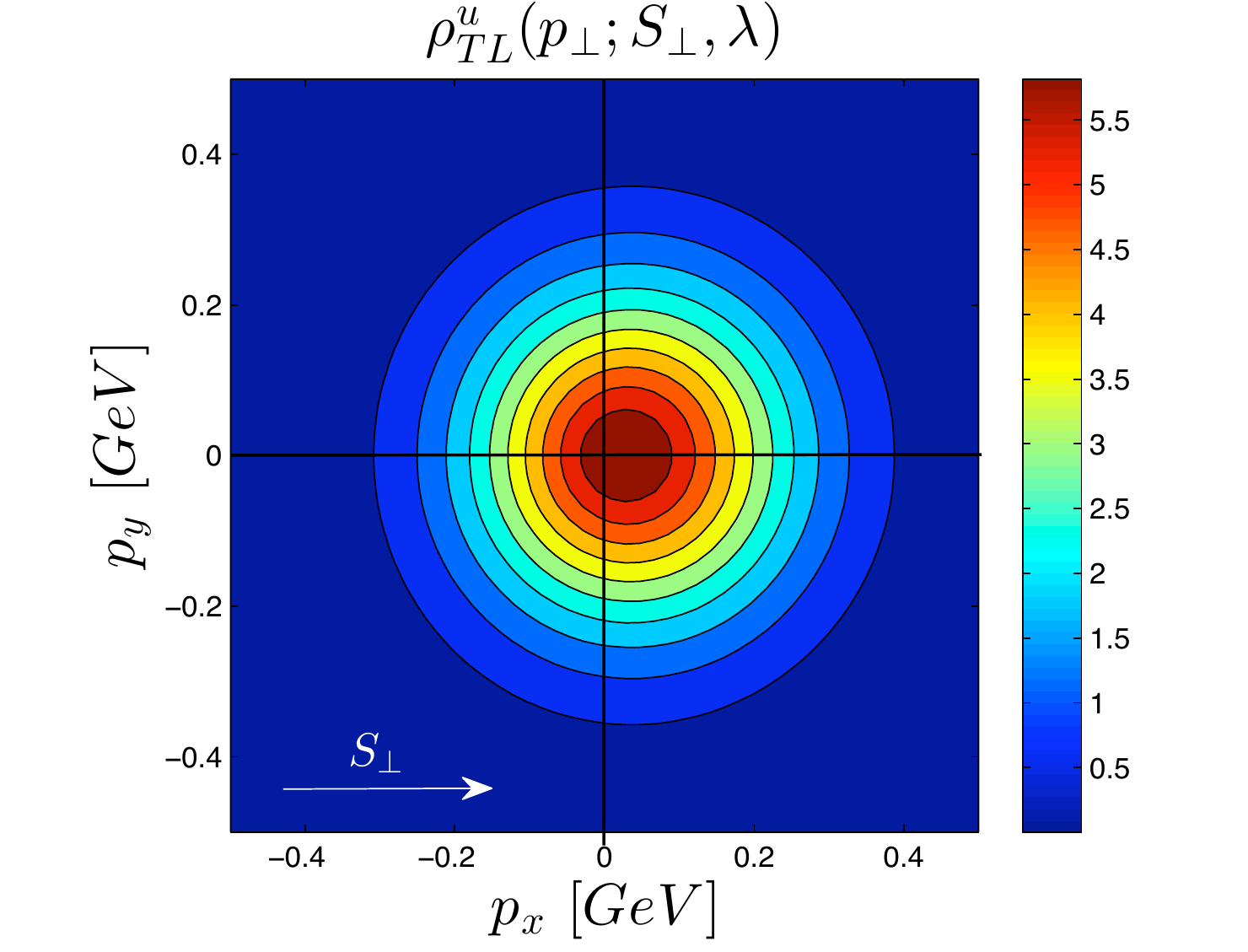}
\hspace{0.1cm}%
\small{(d)}\includegraphics[width=7.5cm,height=5.5cm,clip]{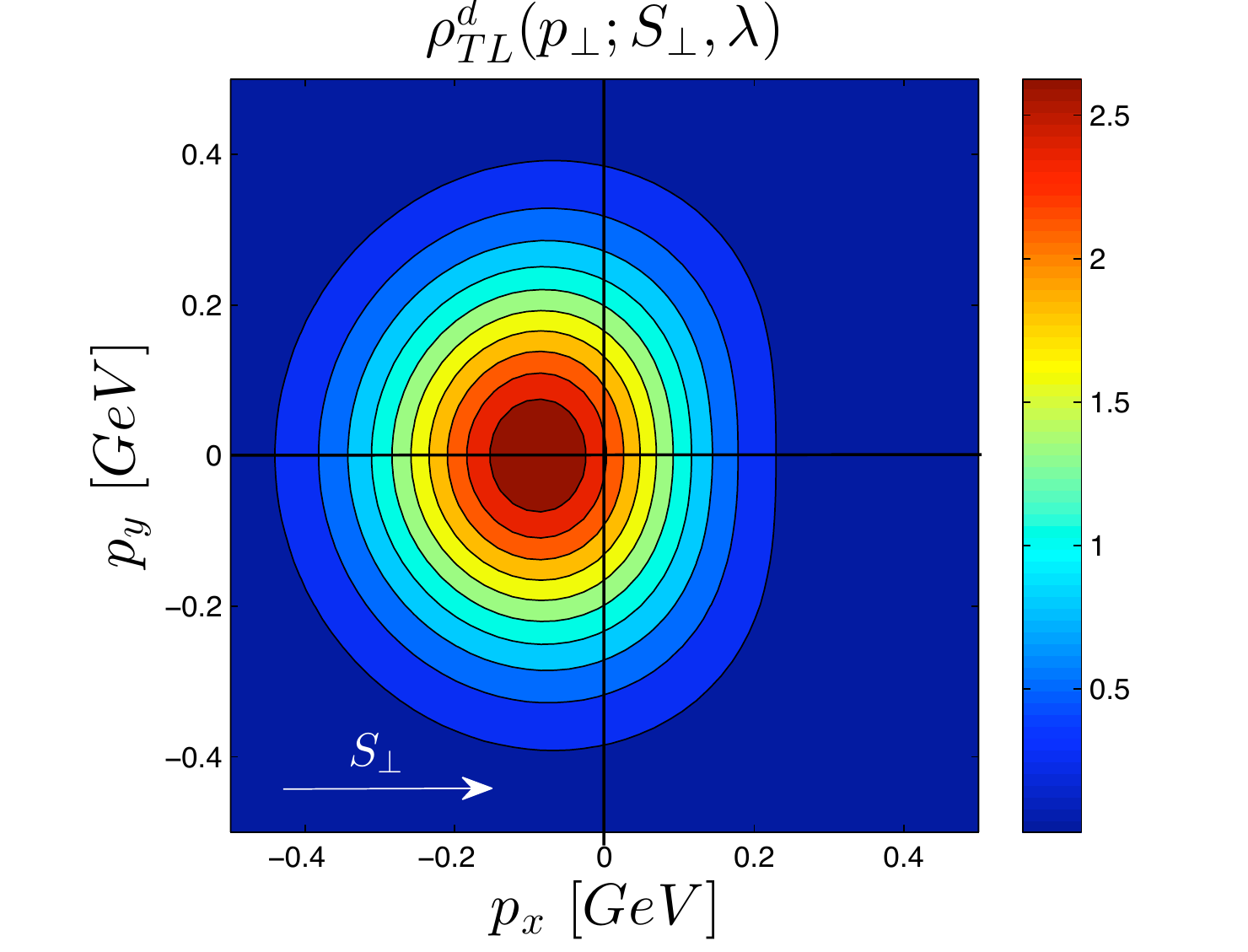}
\end{minipage}
\caption{$\textbf{p}_\perp$-densities of quarks in the two dimensional transverse momentum plane with $\lambda = 1$ and $\textbf{S}_\perp = (1,0)$ for unpolarised and Transversely polarised proton corresponding to u and d quarks\cite{Hagler}. The figure-(c) is for u-d quarks having transversely polarised proton.} \label{fig:rho}
\end{figure*}

The Mellin moments of the TMDs can be interpreted as the densities of quarks inside the nucleon as:
\be
\rho_{UU}(\textbf{p}_\perp) &=& f^{(1)}_1(\textbf{p}^2_\perp),\\ \label{rhoUU}
\rho_{TL}(\textbf{p}_\perp;\textbf{S}_\perp,\lambda) &=& \frac{1}{2} f^{(1)}_1(\textbf{p}^2_\perp)+\frac{\lambda}{2}\frac{\textbf{p}_\perp.\textbf{S}_\perp}{M} g^{(1)}_{1T}(\textbf{p}^2_\perp).\nonumber\\\label{rhoTL}
\ee 
The density $\rho_{UU}(\textbf{p}_\perp)$ is found when  both nucleon and quark  are unpolarised and the $\rho_{TL}(\textbf{p}_\perp)$ is for transversely polarised nucleon. Considering  the spin pointing along
$z$-direction and nucleon polarised in transverse x-direction,$\textbf{S}_\perp = (1,0)$, the $\textbf{p}_\perp$-densities of quarks in the two dimensional transverse momentum plane are plotted in Fig.(\ref{fig:rho}).
 
In our model the unpolarised distributions, the unpolarised nucleon having unpolarised quarks inside, are symmetric for both u and d quarks as predicted from lattice data in \cite{Hagler}. The values of $\langle\textbf{p}^2_\perp\rangle_{\rho_{UU}} = 0.0569 ~GeV^2 $ for u and $ \langle \textbf{p}^2_\perp \rangle_{\rho_{UU}} = 0.0725 ~GeV^2 $ for d quarks.

The quark densities for  unpolarized and polarized proton are shown  in Fig\ref{fig:rho}. Fig.\ref{fig:rho}(a)-(b) represent the densities for $u$ and $d$ quark in unpolarized proton and  Fig.\ref{fig:rho}.(c)-(d), represent the same densities   when proton is transversely polarised along x-axis, $S_\perp =(1,0)$ and the quark spin pointing towards us  with helicity $\lambda=1$. The transverse momentum dependent densities of quarks are no longer axially symmetric in the transverse momentum plane. The peaks sift along the $S_\perp$ but in opposite direction with amplitudes $\langle \textbf{p}_x\rangle_{\rho_{TL}} \approx + 66~ MeV$  for u and $\langle \textbf{p}_x\rangle_{\rho_{TL}} \approx - 85 ~ MeV $ for d quark.
This sifting is because of non-zero $g^{(1)}_{1T}(x,\textbf{p}^2_\perp)$. The deformation in $\rho_{TL}$ indicates that the transversely polarised nucleon is non-spherical, the u and d quarks have opposite directional distributions.

\section{Generalized parton distributions}
Using the overlap formalism of light front wave functions, we evaluate the GPDs in light front quark-diquark model. The  GPDs  $H$ and $E$  are defined through the matrix element of the bilocal vector current on the light-front:
\be
&&\int \frac{dy^-}{8\pi}e^{ixP^+y^-/2}\langle P',\lambda'|~\bar{\psi(0)}\gamma^+\psi(y)~|P, \lambda\rangle = \nonumber\\
&&\!\!\!\!\!\!\frac{1}{2P^+}\bar{U}_{\lambda'}(P')\bigg[H(x,\zeta,t)\gamma^++E(x,\zeta,t)\frac{i}{2M}\sigma^{+\alpha}q_{\alpha}\bigg]U_\lambda(P),\nonumber\\
\label{gpds}
\ee
where $\bar{P}(P')$ is the initial(final) proton momentum, $t=q^2=(P'-P)^2$ and  $\lambda(\lambda^\prime)=\pm\frac{1}{2}$ is the initial (final) proton spin.
In terms of LFWF, the GPDs(for $\zeta=0$) are then obtained  as
\be
H^q(x,t)&=& \int \frac{d^2\bfk}{16\pi^3}~\bigg[\psi_{+q}^{+*}(x,\bfk')\psi_{+q}^+(x,\bfk) \nonumber\\
&+&\psi_{-q}^{+*}(x,\bfk')\psi_{-q}^+(x,\bfk)\bigg],\label{H_gpd}\\
\nonumber\\
E^q(x,t)&=& -\frac{2M}{q^1-iq^2} \int \frac{d^2\bfk}{16\pi^3}~\bigg[\psi_{+q}^{+*}(x,\bfk')\psi_{+q}^-(x,\bfk) \nonumber\\
&+&\psi_{-q}^{+*}(x,\bfk')\psi_{-q}^-(x,\bfk)\bigg],\label{E_gpd}
\ee
where $\bfk'=\bfk+(1-x)\bfq$. Integrating over $k_\perp$, we get
\be
H^q_v(x,t)&\!\!=&\!\!\! \Big[F_1(x)(1-x)^2+F_2(x)(1-x)^4 \frac{\kappa^2}{M^2\log(1/x)}\nonumber\\
\times&&\!\!\!\!\!\!\!\!\!\!\Big(1-\frac{Q^2}{4\kappa^2}\log(1/x)\Big)\Big]\exp \bigg[-\frac{Q^2}{4\kappa^2}\log(1/x)\bigg],\label{H}\\
E^q_v(x,t)&=& 2F_3(x)(1-x)^3 \exp \bigg[-\frac{Q^2}{4\kappa^2}\log(1/x)\bigg].\label{E}
\ee
 The GPDs satisfy the physical conditions
 \be
\int_0^1 dx H^q(x,0,)&=&n_q,\nonumber\\
\int_0^1 dx E^q(x,0,)&=&\kappa_q,
\ee
where $n_q$ denotes the number of $u$ or $d$ valence quarks in the proton and $\kappa_q$ is the  anomalous magnetic moment of quark $q$. 
Note that the exponential factors in both TMDs and GPDs come from the AdS/QCD wavefunction, but due to  additional integration over the transverse momentum, the $t$ dependence in GPD is totally different from the $p_\perp^2$ dependence in the TMDs. 
since $Q^2=-q^2=q_\perp^2=-t$, the exponential factors in Eq.(\ref{H}) and Eq.(\ref{E}), can be written as
$\exp \big[-\frac{Q^2}{4\kappa^2}\log(1/x)\big]=x^{-\alpha^\prime t}$ where $\alpha^\prime =1/(4\kappa^2)$, i.e., the GPDs show Regge behavior. 
But  due to the extra $(1-x)^{-2}$ factor in the exponentials, it is not possible to express TMDs in the Regge type form similar to GPDs.


\subsection{Relations between TMDs and GPDs}
\begin{figure}[h]
\includegraphics[width=6.5cm,clip]{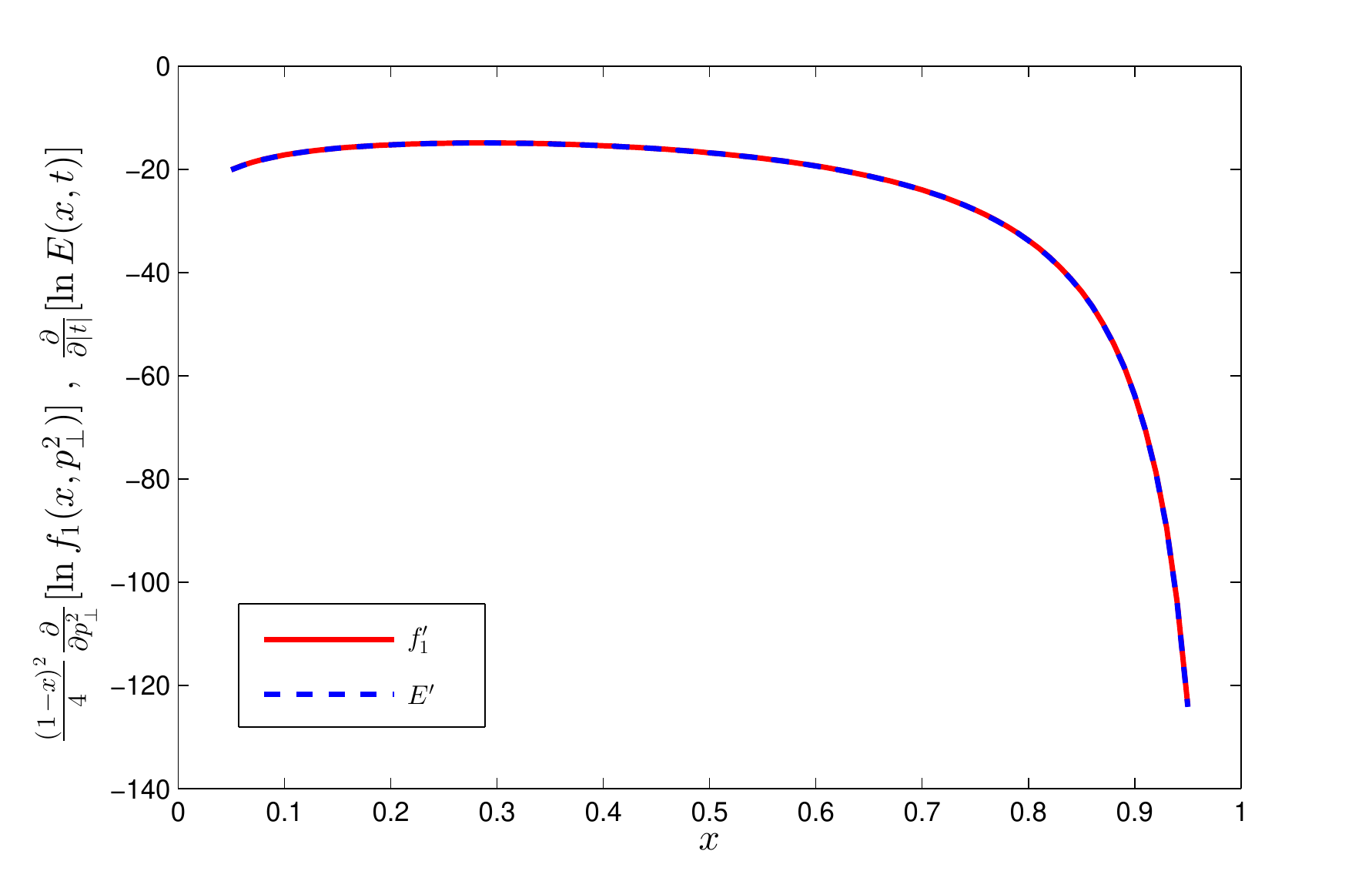} 
 \caption{The variation of $f'_1 = \frac{(1-x)^2}{4}  \frac{\partial}{\partial p^2_\perp}[ \ln f_1(x,p^2_\perp)]$ and   $E'=\frac{\partial}{\partial |t|}[ \ln E(x,t)]$.
\label{log_rel}}
\end{figure}
The GPDs and TMDs satisfy a universal relation:
\be
\frac{\partial}{\partial |t|}[ \ln({\rm GPD})] = \frac{(1-x)^2}{4}\frac{\partial}{\partial p^2_\perp}[ \ln( {\rm TMD})].\label{TMD_GPD}
\ee
This result relies on the particular 
$t$ and $p_\perp^2$ dependence of the GPDs and TMDs  which comes from the AdS/QCD wave function. 
The relation is  not exact for all the TMDs and GPDs, but numerically the differences are found to be insignificant. 

Let us discuss the specific origin of this relation in some detail. The TMDs and GPDs have the two distinct exponential behavior which comes from the exponential behavior of the wavefunction predicted by AdS/QCD, but due to the {\it extra} integration over the transverse momentum transfer, the exponential in GPDs is completely different from that of TMDs. In case of GPDs, the $t $ (or $Q^2$) behavior is of Regge type ($\propto x^{\alpha+\alpha' t}$) but it is rather Gaussian for TMDs.

We expect that this relation, being approximate and model-dependent, may reflect the physics of AdS/QCD duality (which is itself based on {\it approximate} conformal invariance and cannot be exact)
and hold in more general context.

The relation has been demonstrated  for the GPD $E(x,t)$ and TMD $f_1(x,p_\perp^2)$  in Fig.\ref{log_rel}.

The GPDs obtained from the model satisfy the Regge behavior with the Regge slope $\alpha^\prime=1/(4\kappa^2)$.  The average value of Gaussian transverse momentum distribution is then obtained as 
\be
\langle p_\perp^2\rangle=\frac{(1-x)^2}{4 \alpha^\prime \ln{1/x}}=\frac{\kappa^2 (1-x)^2}{ \ln{1/x}},
\ee 
which takes the maximum value of $ \langle p_\perp^2\rangle_{max}\approx (255 \rm{MeV})^2$ for $x \sim 1/3$. 
\begin{figure}[h]
\includegraphics[width=5.5cm,clip]{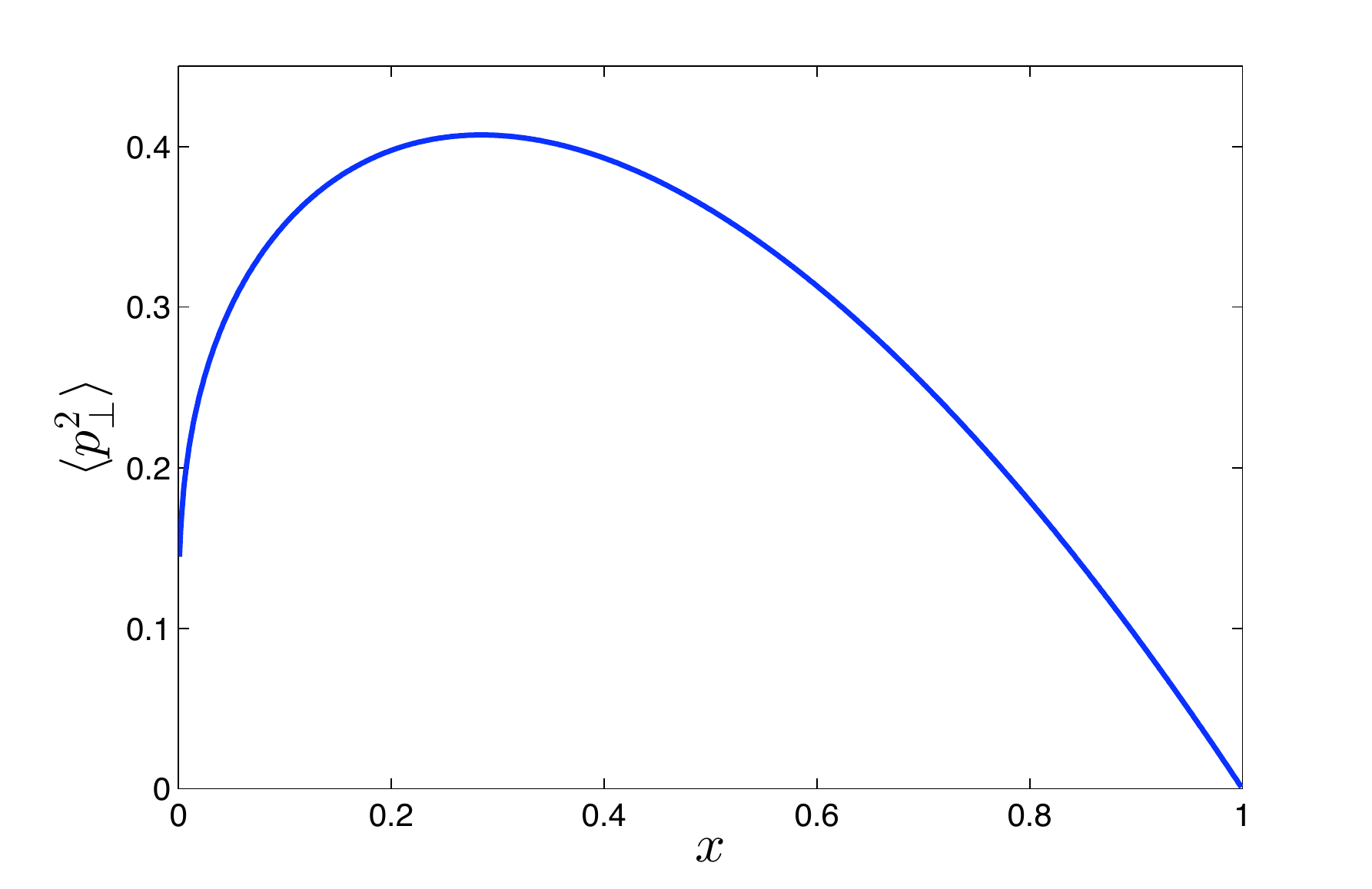}
\caption{The $x-$dependence of average transverse momentum squared in the units of $\kappa^2$ }
\label{figpT2}
\end{figure}
It is interesting that this value of $x$ naturally corresponding to average momentum of valence quark appears here 
as a maximum of elementary function $(1-x)^2/\ln{(1/x)}$ implied by AdS/QCD. 
 This expression also relies on the particular model and possibility to drop the above mentioned numerically 
small terms. Indeed,
such completely different origin of the number naturally explained in the valence quarks picture may seem 
strange and may be accidental, although one cannot exclude that it is yet another manifestation of approximate AdS/QCD duality.

The magnitude of the transverse momentum is independent of quark flavor.  
Note that AdS/QCD warp parameter  $\kappa$ thus corresponds to the seemingly different quantities like
Regge slope and average quark transverse momentum. 
Note also that the Regge slope relation to average transverse momentum was first found by V.N. Gribov in the framework of scalar ladders summation
for soft Pomeron (having some similarity to the case under consideration because of limited transverse momentum) forming a ``Heterotic Pomeron"  \cite{Levin:1992ys} while combining with hard BFKL Pomeron \footnote{O.T. is indebted to 
L.N. Lipatov for pointing out that connection}.

\begin{figure}[h]
\includegraphics[width=7.5cm,clip]{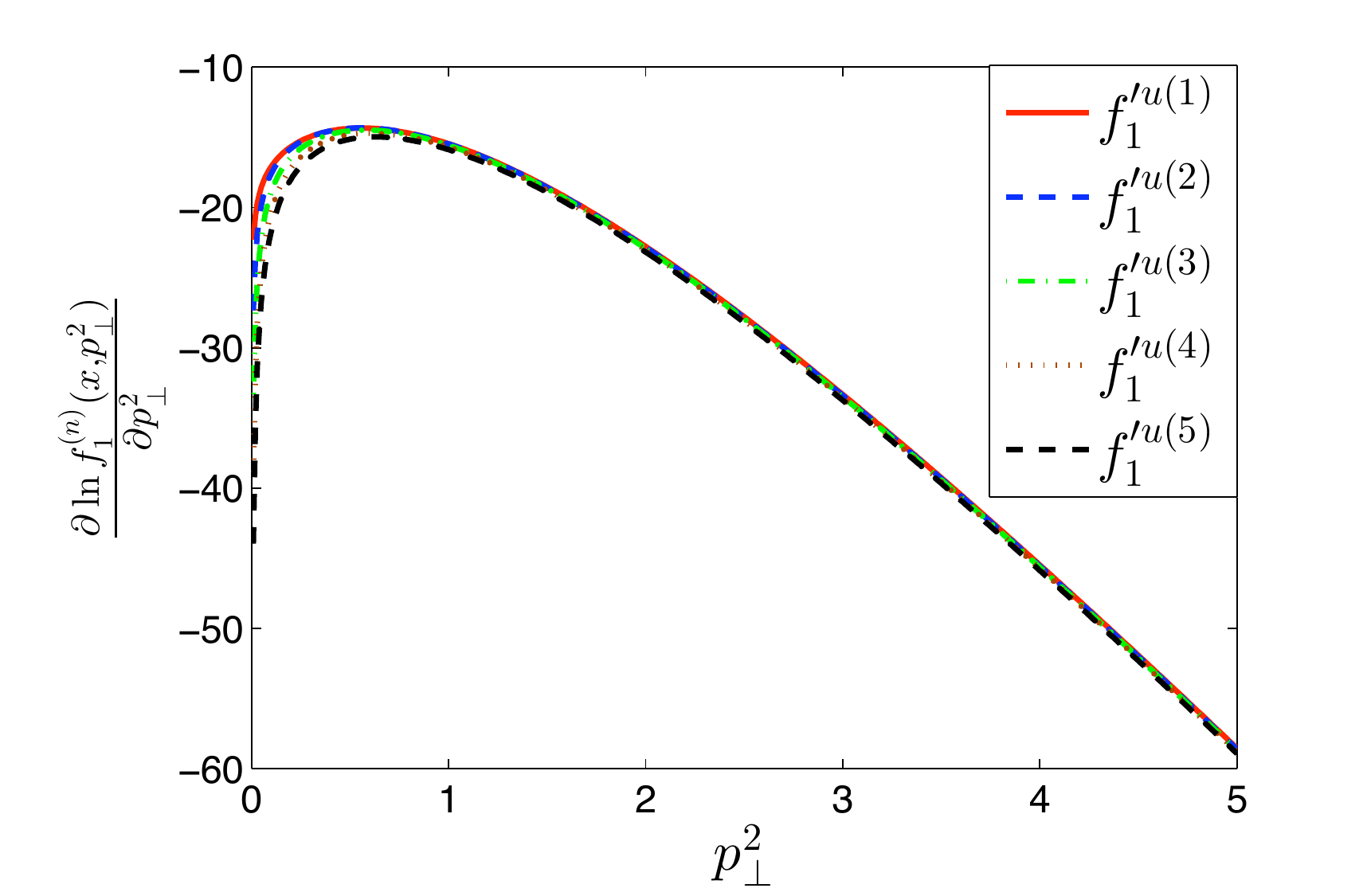}
\caption{\label{f1moments} Illustration of Eq.(\ref{12345mom_f1}).}
\end{figure}
The moments of $f_1(x,p_\perp)$ satisfy another interesting  relation
\be
\frac{\partial}{\partial P^2_\perp}[\ln f^{(1q)}_1(p^2_\perp)] &\simeq& \frac{\partial}{\partial P^2_\perp}[\ln f^{(2q)}_1(p^2_\perp)] \simeq \nonumber\\
\frac{\partial}{\partial P^2_\perp}[\ln f^{(3q)}_1(p^2_\perp)] &\simeq&\cdots \simeq \frac{\partial}{\partial P^2_\perp}[\ln f^{(nq)}_1(p^2_\perp)] \label{12345mom_f1}
\ee
which is illustrated in Fig.\ref{f1moments} for $u$-quarks upto $5$-th moments. We obtain similar plot for $d$-quark also.
 
\begin{figure*}[h]
\begin{minipage}[c]{0.98\textwidth}
\small{(a})\includegraphics[width=7.5cm,clip]{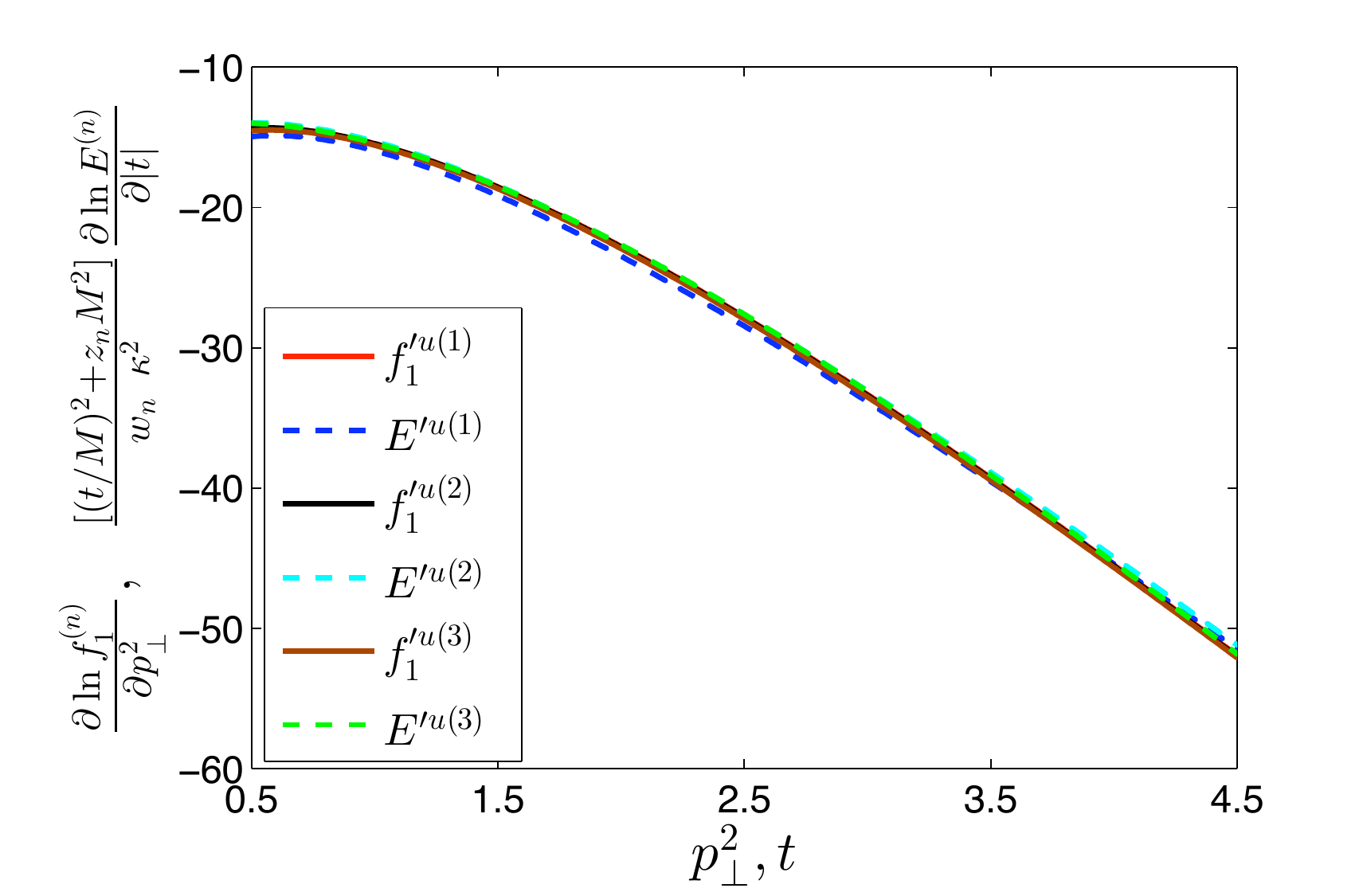}
\hspace{0.1cm}%
\small{(b)}\includegraphics[width=7.5cm,clip]{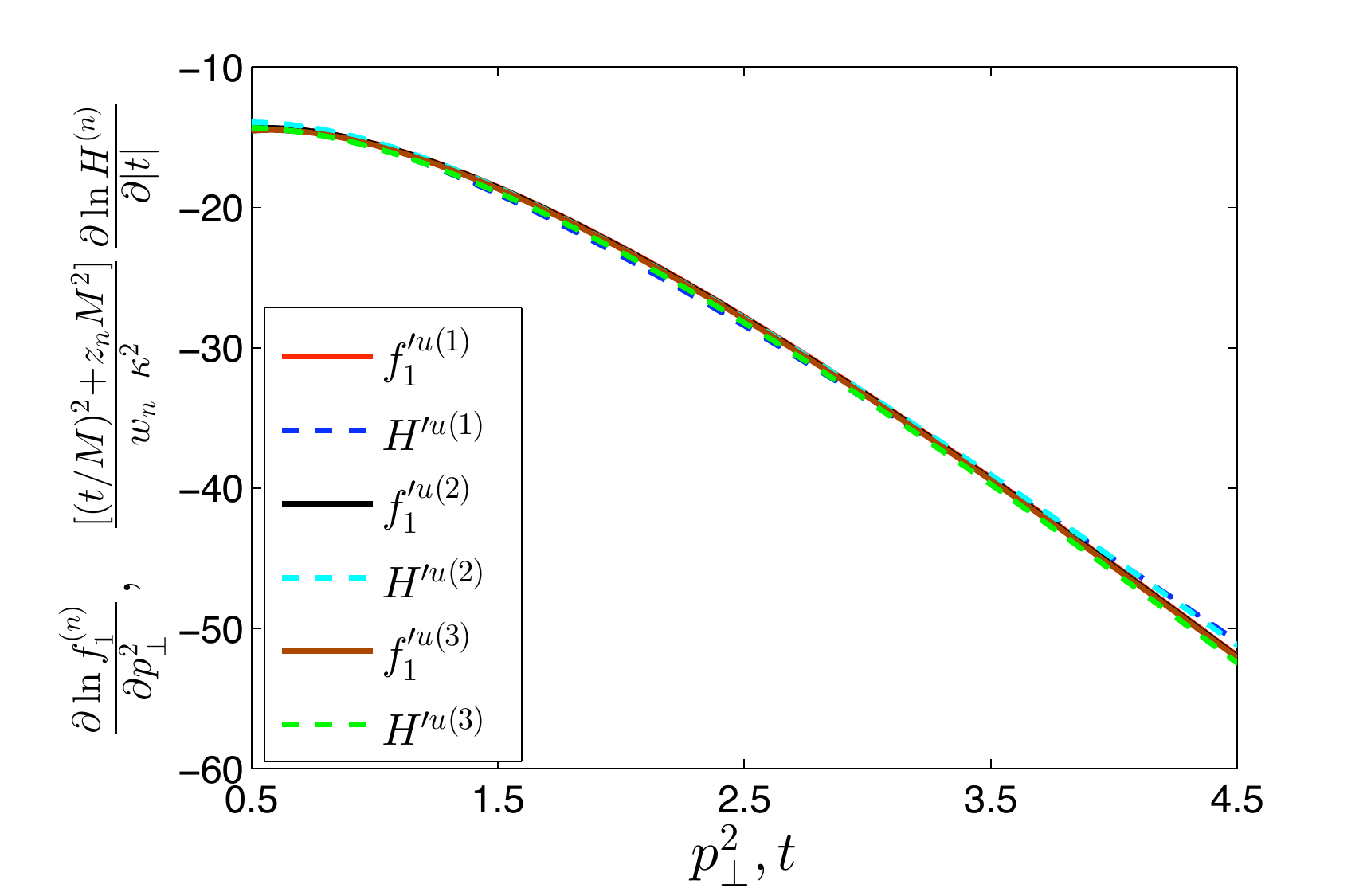}
\end{minipage}
\begin{minipage}[c]{0.98\textwidth}
\small{(c)}\includegraphics[width=7.5cm,clip]{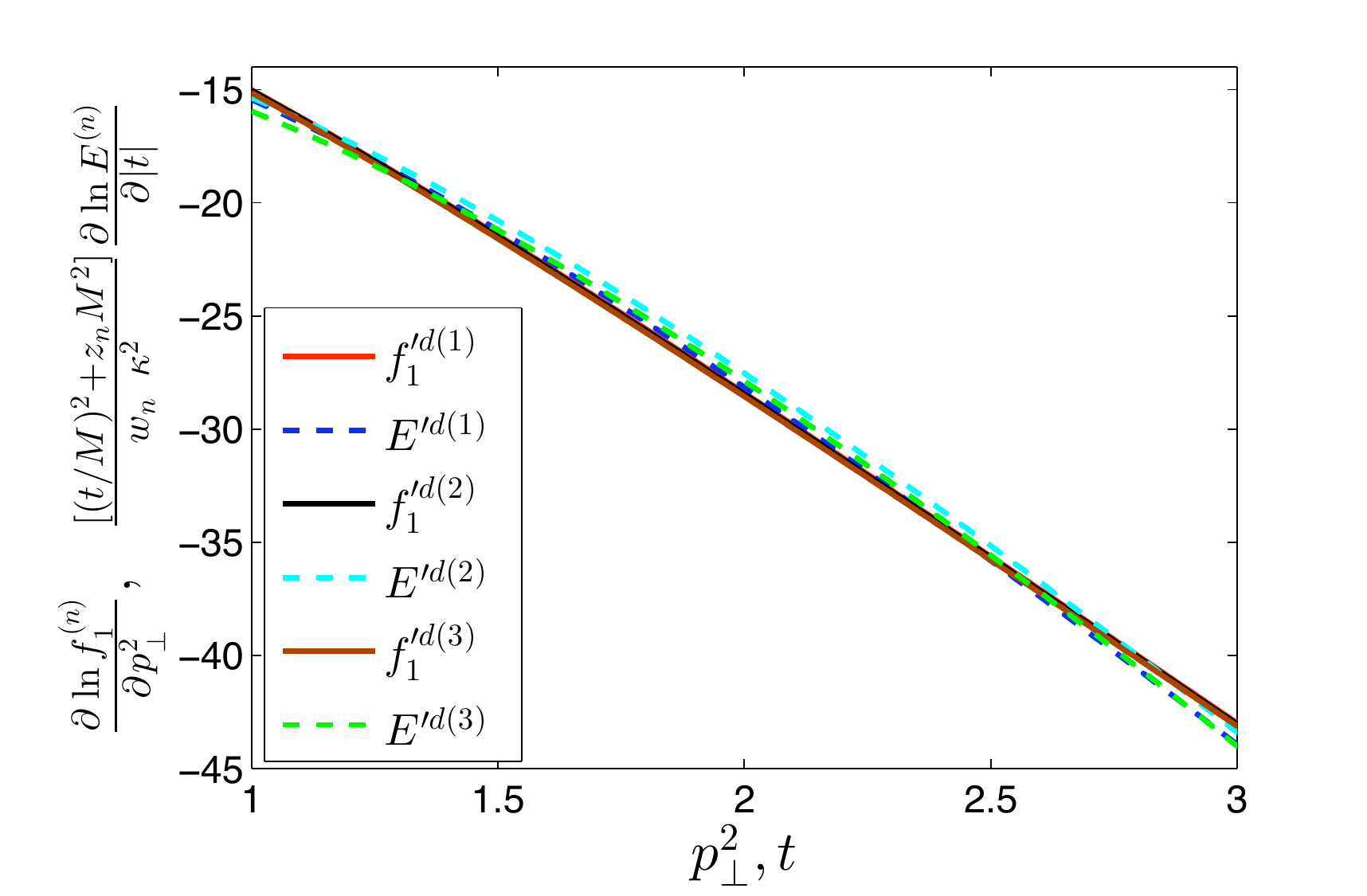}
\hspace{0.1cm}%
\small{(b)}\includegraphics[width=7.5cm,clip]{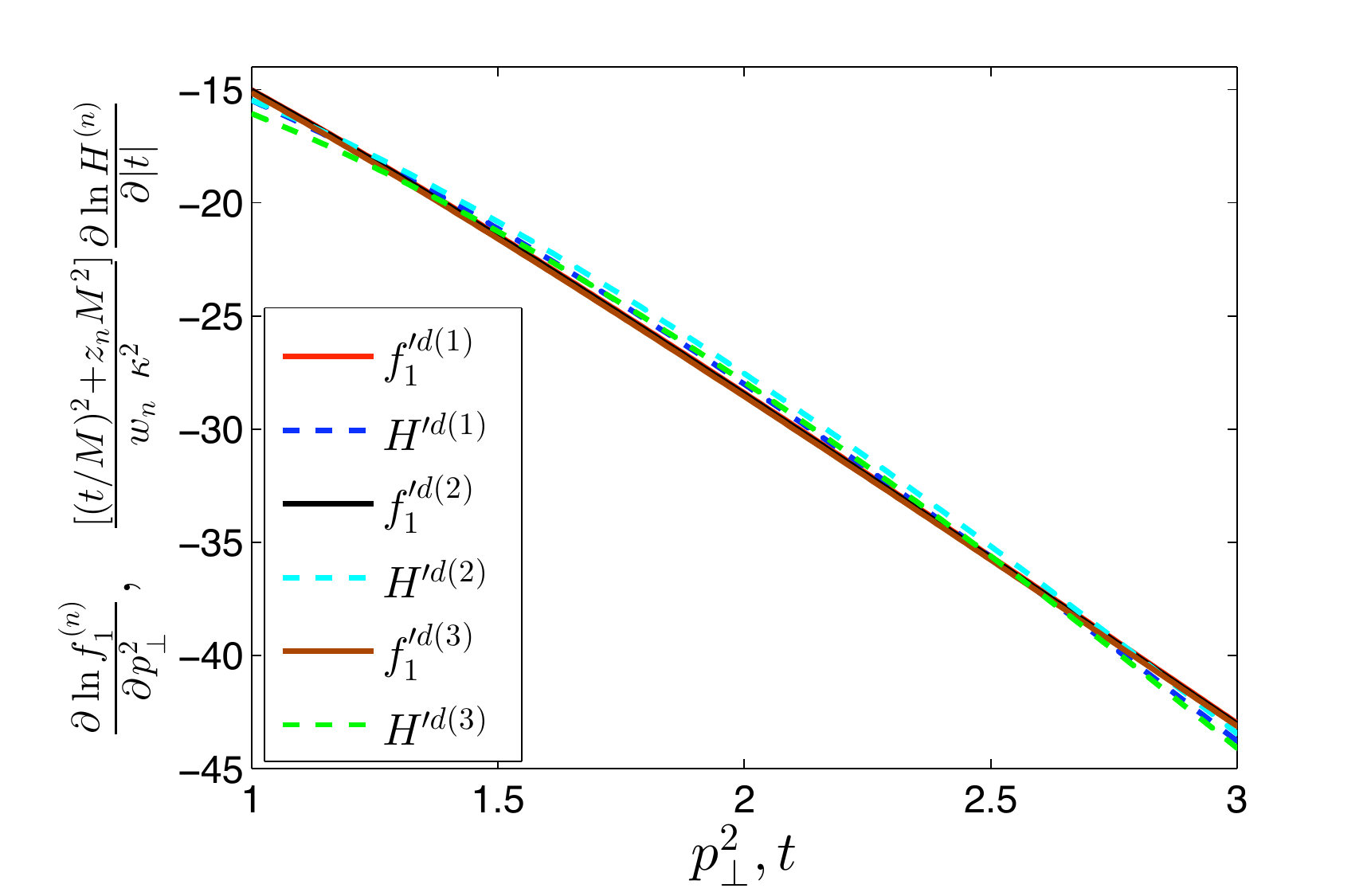}
\end{minipage}
\caption{\label{empirical1} Plot of Eq.(\ref{momRel}) for$f_1(x,p_\perp)$ and (a)  $E_u(x,t)$, (b) $H_u(x,t)$, (c)$E_d(x,t)$ and (d) $H_d(x,t)$ for $n=1,2,3$.}
\end{figure*}

 In the high momentum region 
 the slopes of the $n$th moments of TMDs and GPDs satisfy the following  empirical but a relation general to all  $n$ (numerically checked upto $n=5$):
\begin{small}
\be 
\!\!\!\!\!\!\frac{\partial}{\partial |t|}\ln({\rm GPD})^{(n)}(t) \simeq  \frac{w^q_n\kappa^2}{[(\frac{t}{M})^2+z^q_n M^2]}\frac{\partial}{\partial P^2_\perp}\ln f^{(n)}_1(p^2_\perp), \label{momRel}
\ee
\end{small}
where the superscript $(n)$ stands for the $n$-th moment.  $z^q_n$ and $w^q_n$  are given by
 $z^u_n = 2+0.5(n-1),~ w^u_n = 1.5-0.12(n-1)$ when the GPD is  $H_u(x,t)$ and $z^u_n=1.7+(n-1)0.55,~w^u_n=  1.5-(n-1)0.12$ for $E_u(x,t)$ and for the d-quark the values are
 $z^d_n= 1.06+(n-1)0.385,~w^d_n =1.15-(n-1)0.105$ for $H_d$ and $z^d_n= 1.7+(n-1)0.55,~w^d_n=1.12-(n-1)0.1$ for $E_d$.
These  relations  for $u$ and $d$ quarks are illustrated in Figs.\ref{empirical1}  for first three moments.
\begin{figure*}[h]
\begin{minipage}[c]{0.98\textwidth}
\small{(a)}
\includegraphics[width=7.5cm,clip]{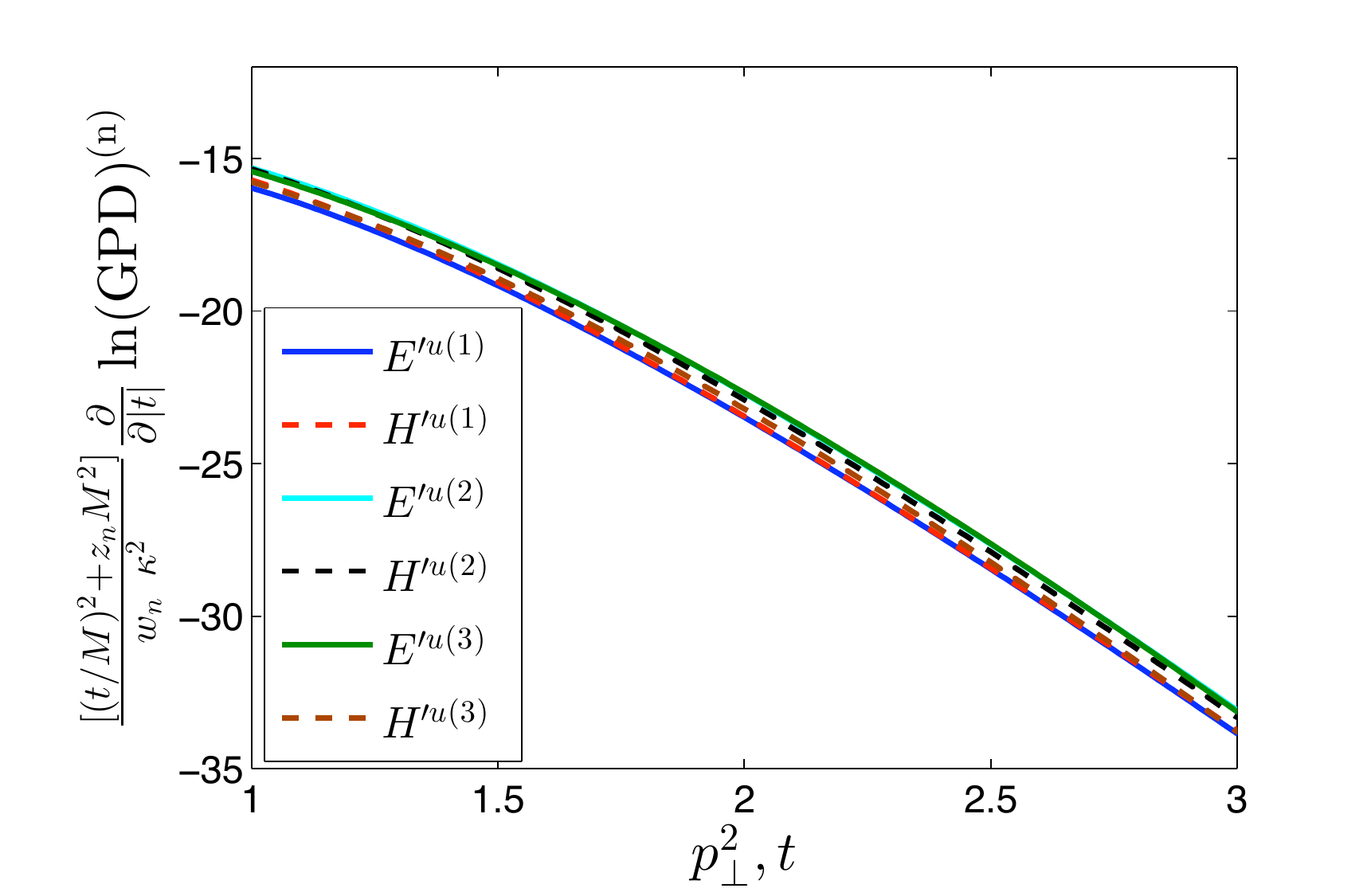}
\hspace{0.1cm}%
\small{(b)}\includegraphics[width=7.5cm,clip]{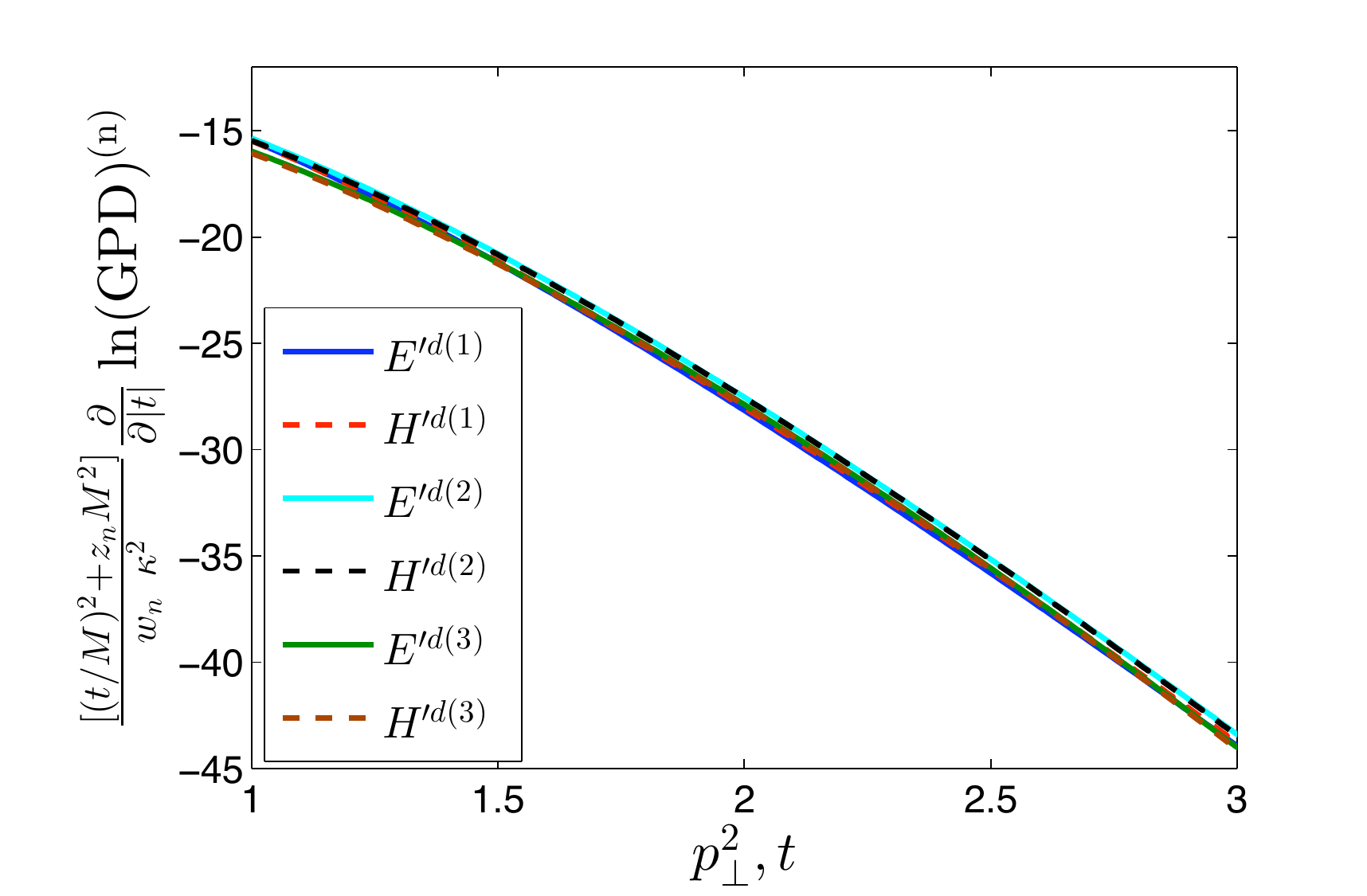}
\end{minipage}
\caption{  The Eq.(\ref{logRel_EGPD_HGPD})is plotted  with $n=1,2,3$ for (a) u quark and (b) d quark. \label{logRel_GPDs}}
\end{figure*}
This difference of transverse momentum dependence may be qualitatively explained by the different x-dependence of exponential 
factors in the corresponding integrand. As for GPDs it is just $\exp(-|t \alpha^\prime \ln{x}|)$, for larger $t$ the larger values of $x$ contribute so that 
the decrease of the moments is only power-like, the notable example represented by electromagnetic form factors. At the same time, for TMDs 
the similar factor is $\exp(-|p_\perp^2 \alpha^\prime \ln{x}/4(1-x)^2 |)$ so that it is suppressing the contributions of both large and small $x$ and 
only the region of $x \sim 1/3$ corresponding to the maximum at Fig.\ref{figpT2} contributes. As a result, while using the saddle-point method, the corresponding exponential factor may be taken out of the integral and the main dependence of TMD moments on $p_\perp^2$ is an {\it exponential} one! This explains the extra factor in 
Eq.(\ref{momRel}).
This also may be the the reason of  approximate  ``factorization" of $x$ and $p_\perp$ dependence of TMDs 
which is not contradicting to 
lattice QCD  data \cite{Hagler,lattice}. Such factorization ansatz is also  
 used in the phenomenological parameterizations and experimental extractions of the TMDs\cite{anselmino}.
\begin{figure*}[h]
\begin{minipage}[c]{0.98\textwidth}
\small{(a})\includegraphics[width=7.5cm,clip]{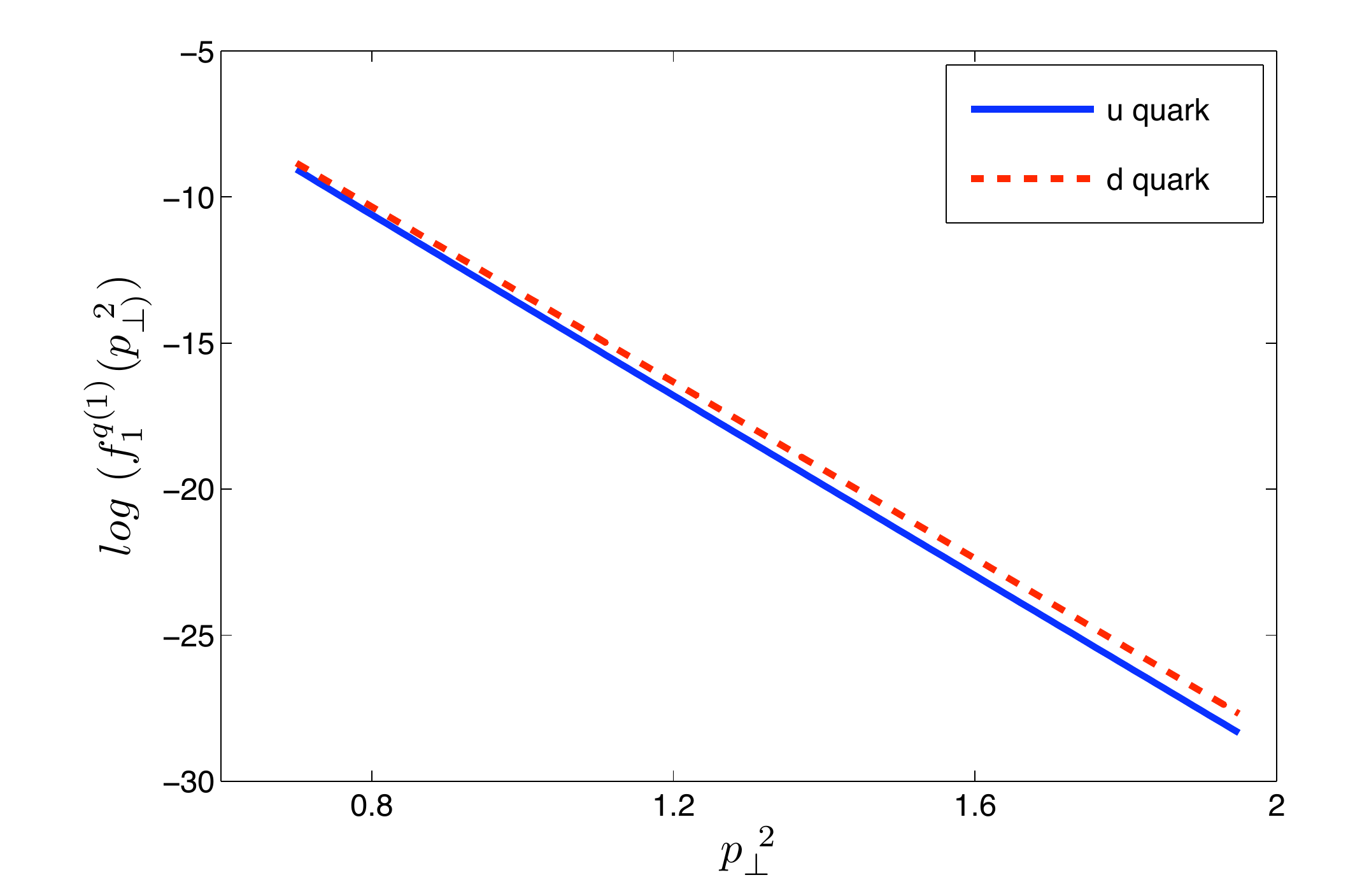}
\hspace{0.1cm}%
\small{(b)}\includegraphics[width=7.5cm,clip]{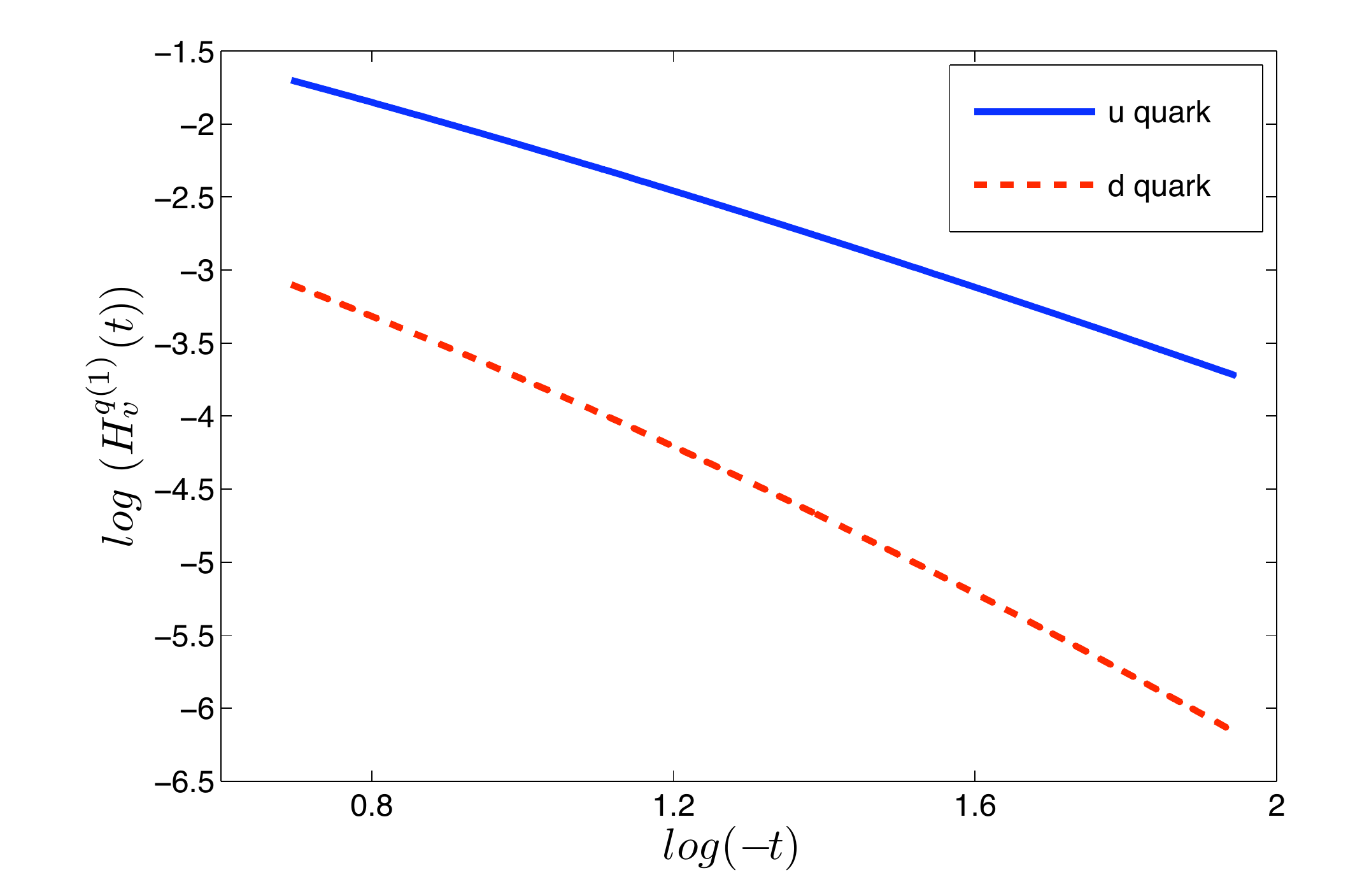}
\end{minipage}
\caption{\label{1st_moments} (a) The $p_\perp^2$ dependence of the first moments of the  TMD $f_1^q(x,p_\perp^2)$ and (b) $t$-dependence of the first moment of the GPD $H_v^q(x,t)$}
\end{figure*}

 It is interesting to note that with the same values of $z_n$ and $w_n$ 
one can relate the slopes of the two GPDs also
\be
&&\frac{[(\frac{t}{M})^2+z^{q,E}_n M^2]}{w^{q,E}_n\kappa^2}\frac{\partial}{\partial |t|}[\ln\rm{E}^{(n)}(t)] \simeq  \nonumber\\
&&~~~\frac{[(\frac{t}{M})^2+z^{q,H}_n M^2]}{w^{q,H}_n\kappa^2}\frac{\partial}{\partial |t|} [\ln\rm{H}^{(n)}(t)], \label{logRel_EGPD_HGPD}
\ee
as can be seen in Fig.\ref{logRel_GPDs}. Since, the moments of different distributions are calculate in lattice QCD, these relations can help to check the consistency of model calculations with lattice results.

To provide some quantitative ground for this picture we calculated the transverse momentum dependence of the moments of TMDs and GPDs. 
In Fig.\ref{1st_moments} (a), we have shown the first moment of $f_1^q(x,p_\perp^2)$. It clearly indicates that the first moment of the TMD has exponential dependence on the transverse momentum $p_\perp^2$. This behavior of the moments is a necessary condition of the mentioned factorization of $x$ and $p_\perp$ dependencies,
At the same time,  Fig.\ref{1st_moments}(b), indicates that for large $-t$, the GPD follows a power law behavior with $-t$ which is naturally related to the Regge Parameterization.

\section{Discussion} 

We calculated the TMDs in the framework of soft wall AdS/QCD based diquark model of the nucleon and performed the systematic 
exploration of their properties and comparison with GPDs. We tested the number of relations  and inequalities for TMDs.

The fit of data leads to the value of warp parameter $\kappa \approx 400MeV$.

The new relation between $t$-dependence of GPDs and $p_\perp$-dependence of TMDs is found.
 It is approximate and model dependent, but may express the physics of approximate AdS/QCD duality, 
which requires further checks.
The x dependence of average transverse momentum squared or TMDs is calculated.  
While the GPDs exhibit the Regge behavior combining the exponential dependence on $t$ at fixed $x$ with the power dependence of the moments, 
related to the formfactors, the TMDs exhibit the gaussian dependence on $p_\perp$ both at fixed $x$ and after integration over $x$.
This may be the  reason for approximate factorization of $x$ and $p_\perp$ dependencies of TMDs compatible with lattice QCD and used 
in phenomenological analysis.

 We thank A. V. Efremov, L.N. Lipatov and C.~Lorc\'e for useful discussions and comments.  
 D.C. is grateful to BLTP, JINR, where this work was completed, for warm hospitality.
 O.T. was supported in part by RFBR grant  14-01-00647.


\end{document}